# Phase field theory for fracture at large strains including surface stresses

**Hossein Jafarzadeh[a], Gholam Hossein Farrahi[a,*], Valery I. Levitas[b,c], Mahdi Javanbakht[d]**

[a] Sharif University of Technology, School of Mechanical Engineering, Tehran 11365-11155, Iran

[b] Iowa State University, Departments of Aerospace Engineering and Mechanical Engineering, Ames, IA 50011, USA

[c] Ames Laboratory, Division of Materials Science and Engineering, Ames, IA, USA

[d] Isfahan University of Technology, Department of Mechanical Engineering, Isfahan 84156-83111, Iran

**Abstract**

Phase field theory for fracture is developed at large strains with an emphasis on a correct introduction of surface stresses. This is achieved by multiplying the cohesion and gradient energies by the local ratio of the crack surface areas in the deformed and undeformed configurations and with the gradient energy in terms of the gradient of the order parameter in the reference configuration. This results in an expression for the surface stresses which is consistent with the sharp surface approach. Namely, the structural part of the Cauchy surface stress represents an isotropic biaxial tension, with the magnitude of a force per unit length equal to the surface energy. The surface stresses are a result of the geometric nonlinearities, even when strains are infinitesimal. They make multiple contributions to the Ginzburg-Landau equation for damage evolution, both in the deformed and undeformed configurations. Important connections between material parameters are obtained using an analytical solution for two separating surfaces, as well as an analysis of the stress-strain curves for homogeneous tension for different degradation and interpolation functions. A complete system of equations is presented in the undeformed and deformed configurations. All the phase field parameters are obtained utilizing the existing first principle simulations for the uniaxial tension of Si crystal in the [100] and [111] directions.

1. Introduction

*Phase field method for fracture.* The Ginzburg-Landau or the phase field method is a powerful approach for simulation of complex microstructures. The phase field approaches have some advantages. In particular, they provide the possibility of describing the evolution of an arbitrary and complex evolving crack geometry without requiring a priori information or additional computational efforts to track crack paths. Interaction with discrete (precipitates, different interfaces, and inclusions) and continuous heterogeneities does not require additional computational efforts either. A phase field model is commonly associated with an order parameter. There are different definitions for the order parameter, depending on the discipline and the purpose. We refer to the order parameter as a thermodynamic variable which describes some type of material instability/stability during microstructure evolutions such as fracture (Amor, Marigo, & Maurini, 2009; Bourdin, Larsen, & Richardson, 2011; Farrahi, Javanbakht, & Jafarzadeh, 2018; Hakim and Karma, 2009; Henry and Levine, 2004; Karma and Lobkovsky, 2004; Kuhn and Müller, 2010; Levitas, Idesman, & Palakala, 2011; Levitas, Jafarzadeh, Farrahi, & Javanbakht, 2018; Miehe, Aldakheel, & Raina, 2016; Miehe and Schänzel, 2014; Wang, Jin, & Khachaturyan, 2002; Weinberg and Hesch, 2017), martensitic phase transformations (Levitas, 2013a, b; Levitas, 2014; Levitas and Javanbakht, 2010; Levitas and Warren, 2016), damage (Mozaffari and Voyiadjis, 2015; Mozaffari and Voyiadjis, 2016), etc. For the martensitic phase transformations, the order parameter is related to one of the following: (a) concentrations of martensitic variants, (b) transformation strain tensor, (c) some components of the total strain tensor, or (d) atomic shuffles (intracellular displacements). Fracture is also associated with the displacement of some atoms, similar to the phase transformation. However, in the context of fracture, the displacement of atomic planes leads to atomic bond breaking. Thus, an order parameter $\phi$ is employed to describe the stability of the position of atomic planes during the separation. The intact material, which is the solid state, corresponds to $\phi=0$; the completely broken (damaged) state has $\phi=1$; and within each crack surface, which has a thin thickness, the order parameter continuously varies from 1 to 0. The order parameter $\phi$ describes the bond breaking, and to keep this feature, a single-well potential (Bourdin et al., 2011; Kuhn and Müller, 2010; Levitas et al., 2011; Levitas et al., 2018; Miehe et al., 2016; Wang et al., 2002) is required rather than the double-well one, which treats the crack propagation as a phase transformation from solid to gas (Farrahi et al., 2018; Henry and Levine, 2004; Jafarzadeh, Farrahi, & Javanbakht, 2019; Karma and Lobkovsky, 2004). Note that, in general, the phase field approach to fracture is quite similar to the cohesive zone approach, but with a more advanced and flexible kinetics that does not require a priori knowledge of the crack direction and remeshing.

*Surface stresses within the phase field approach.* The thickness of external surfaces and crack tip radius are of the order of magnitude of few nanometers. Surface tension, been found to play significant roles in the determination of mechanical properties of nanosized materials and structures (Li and Mi, 2019). Thus, the surface stresses should play an essential role in nanoscale simulations of nucleation and propagation of cracks. It is well known that isotropic biaxial stresses with force per unit length *T* (Porter, Easterling, & Sherif, 2009) act on each material interface or surface. For liquid-liquid and liquid-gas interfaces, the surface stresses are independent of deformation because they do not support elastic stresses. The force per unit length for these interfaces is equal to the surface energy, $T=\gamma$. This leads to a jump in the normal stresses across the interface with the magnitude of $2\gamma/r$, where $r$ is the mean interface radius. However, for a solid surface or interface, the surface stresses have the deformation-dependent part, which can be either tensile or compressive and is related to the surface elasticity.

The Griffith criterion for the crack propagation is based on the surface energy (Griffith, 1921) and does not account for the surface stresses. However, sharp corners and a sharp crack tip are subjected to large surface stresses. Since it is difficult to measure the surface stresses, there is not a lot of quantitative predictions for the effect of the surface stresses on the fracture. In the sharp surface approach, the surface stress tensor (with a dimension of force per unit length, not area) is related to the surface energy by the equation $\bar{\boldsymbol{\sigma}}_s = \gamma \boldsymbol{I}_s + \partial \gamma / \partial \boldsymbol{\varepsilon}_s$, where $\boldsymbol{\varepsilon}_s$ is the surface strain tensor and $\boldsymbol{I}_s$ is a two-dimensional surface unit tensor (Cammarata and Sieradzki, 1994). This equation is usually used (Hu, Lee, & Li, 2018; Li and Wang, 2015; Ou, Wang, & Wang, 2008; Wang and Li, 2013) to consider the effect of the surface stresses on fracture behavior using the linear constitutive law $\partial \gamma / \partial \boldsymbol{\varepsilon}_s = \boldsymbol{C}_s : \boldsymbol{\varepsilon}_s$, where $\boldsymbol{C}_s$ is the surface elastic moduli tensor. The first part of the surface stresses $\gamma \boldsymbol{I}_s$ is the structural part which is similar to that in liquids and gases and the second term $\partial \gamma / \partial \boldsymbol{\varepsilon}_s$ is the strain-dependent part of the surface stresses. The first (elastic) part of the surface stresses can be neglected in the small strain theory (Ou et al., 2008; Wang and Li, 2013) because it is shown in atomistic simulations that the components of $\boldsymbol{C}_s$ are of the same order of magnitude as $\gamma$ (Li and Wang, 2015). Furthermore, the material parameters for constituting the surface stresses are not known well. Another problem is uncertainty as to whether strong heterogeneity across the surface fields of properties, strains, and stresses can be formalized in terms of the resultant stresses without the moments (Levitas, 2014). However, in the phase field approach, the elastic part of the surface stresses comes directly from the coupled solution of the Ginzburg-Landau and elasticity equations. Thus, the elastic stresses localized inside the diffuse (i.e., finite-width) surface present and consider the variation of elastic properties across the surface, the finite surface width, and the heterogeneity of stresses across and along the surface. This includes a description of the strain-dependent surface stresses with much more details than any sharp surface model. Therefore, we only need to include the structural contribution to the surface stresses, which is one of our goals in this paper. Such a problem formulation was suggested in Levitas and Javanbakht (2010), Levitas (2013b), and Levitas (2014) for the martensitic phase transformations. Thus, we will focus on the structural contribution to the surface stresses only.

Wheeler and McFadden, (1997) suggested a general treatment of the interfacial stresses for anisotropic diffuse phase interfaces (including anisotropic interface energy and tension). They utilized the total energy per unit current volume and the gradient of the order parameter in the deformed state. Such assumptions and application of the principle of least action (or Noether's theorem) resulted in an automatic appearance of the interfacial stresses. Similar models, but coupled to mechanics, were developed in Anderson, McFadden, & Wheeler (2001) and Lowengrub and Truskinovsky (1998). As it was shown in Levitas (2013b) and Levitas (2014), stresses obtained in these works were correct for the thermodynamic equilibrium condition and isotropic interfaces. However, they contained an additional hydrostatic pressure in the bulk material for propagating interfaces; this is contradictory because the stresses were not localized at the interface. Note that Hakim and Karma (2009) applied the Noether's theorem-based approach to fracture, in which the energy and the gradient of the order parameter were determined in the reference configuration. Such a formulation did not lead to any surface stress, highlighting the necessity of the utilization of the current configuration for such approaches. The most advanced model for the interfacial stresses during phase transformations is developed in Levitas (2013b) and Levitas and Javanbakht (2010) for small strains and in Levitas (2014) and Levitas and Warren (2016) for large strains. The approach in Levitas (2013b), Levitas (2014), Levitas and Javanbakht (2010), and Levitas and Warren (2016) utilizes the gradient of the order parameter in the current configuration, and the gradient and the double-well energy are defined per unit current volume. A detailed literature review and a comparison of different approaches for the introduction of the interfacial stresses for phase interfaces can be found in Levitas (2014) and Levitas and Warren (2016).

The only phase field approaches to fracture that include the surface stresses were recently developed in Levitas et al. (2018) and Jafarzadeh et al. (2019) for small strains. The approach in Jafarzadeh et al. (2019) is a direct application of the approach for phase transformations from Levitas (2013b) to fracture. However, it was shown in Levitas et al. (2018) that the approaches to the interface stresses developed for phase transformations and based on energies per unit volume of the deformed configuration could not be applied to the fracture problem. This is because, within such an approach, a space between the crack surfaces also possesses a cohesion energy, which violates an energy balance; thus, a new approach is required. Such an approach was developed in Levitas et al. (2018). It also includes geometric nonlinearities even within the small strain formulation: the cohesion and the gradient energies are determined per unit volume in the reference configuration but are multiplied by the ratio of the current to the initial crack surface area, $dS/dS_0$. In the reference configuration, the space between the crack surfaces does not appear and is not energetically penalized, resolving the contradiction above. At the same time, a thermodynamic treatment of the potential with the ratio of the current to the initial crack surface area results in the desired expression for the structural part of the surface stresses. The general theory in Levitas et al. (2018) is illustrated by the finite element solutions of some model problems.

In summary, we are not aware of any phase field model which includes the surface stresses during fracture at large strains. As it was mentioned, the surface stresses are incorporated in Levitas et al. (2018) by introducing some geometric nonlinearities. Thus, a strict treatment of the surface stresses requires a thermodynamically consistent finite strain formulation. Also, at the nanoscale, the material is exposed to large strain before and during fracture. Some of the recently developed phase field models have incorporated a large strain formulation for fracture (Borden et al., 2016; Miehe et al., 2016; Miehe and Schänzel, 2014; Weinberg and Hesch, 2017). However, these papers do not include the surface stresses and have some drawbacks, which will be discussed below.

*Goals and outlook.* Our goal in this paper is to develop a general thermodynamically consistent large-strain phase field approach to fracture which includes the surface stresses. We will use the main ideas for including the surface stresses from Levitas et al. (2018) for the small-strain formulation and will advance and incorporate them into a large-strain approach. Thus, the relationship between the current model and the one in Levitas et al. (2018) is similar to the relationship between small (Levitas, 2013b) and large (Levitas, 2014) strain formulations for the martensitic phase transformations with the interfacial stresses.

Below is the main content of the paper. In Section 2, the integral laws of thermodynamics are presented and localized in the undeformed configuration. A generalized thermodynamic surface force which is conjugated to the order parameter is introduced at the external surface. This allows a stricter treatment of the gradient-type materials. Then, the expression for stresses and the driving force for the evolution of the order parameter for damage are derived. In Section 3, boundary conditions for the order parameter are presented in both the undeformed and deformed states. The structure of the free energy is suggested in Section 4, which leads to the correct expression for the surface stresses. Thus, two terms which determine the surface energy, the cohesive and the isotropic gradient energies, are multiplied by the ratio of the current to the initial crack surface area, $dS/dS_0$. This defines the surface energy per unit current area, similar to the sharp surface approach with the surface tension (Porter et al., 2009). It is shown in Section 5, where the corresponding expressions for the first Piola-Kirchhoff and Cauchy stress tensors are derived, that each consists of the elastic and structural parts; the structural part appears due to the multiplier $dS/dS_0$ in the expression for energy. It is shown that the structural part represents a biaxial tension which its magnitude is equal to the surface energy, reproducing a proper expression for the surface stresses. Detailed expressions for the Ginzburg-Landau equation for the evolution of the order parameter are derived in Section 6. Remarkably, the elastic and surface stresses both explicitly contribute to the evolution equation. In Section 7, a new family of interpolation functions for the cohesive energy is introduced so that the damage starts at finite strains with a significant jump in elastic moduli. An additional requirement for the interpolation function is introduced to ensure a finite width of the damage zone within the crack surfaces. Corresponding conditions and the interpolation function are found. A flexible degradation function with a new parameter $n$, which is used to calibrate the shape of the stress-strain curve for homogeneous tension, is also introduced. This allows for an improved description of the local stress-stress curve at the nanoscale when it is known from the experiment or atomistic simulations. Equilibrium stress-strain curves for any pair of work conjugates are shown in Section 8. In Section 9, the stationary Ginzburg-Landau equation is solved for a static crack for the chosen interpolation function. As expected, the equality of the gradient and the cohesive energies at each point of the surface is shown and used in this section. Section 10 includes an analytical expression of the surface stresses for the current model. A complete system of equations is formulated in Section 11. All the phase field parameters are obtained utilizing the existing first principle simulation results for an uniaxial tension of Si crystal in the [100] and [111] directions in Section 12. Section 13 contains the concluding remarks and future outlooks.

Multiplication and the inner product of two second-order tensors $\boldsymbol{A} = \{A_{ij}\}$ and $\boldsymbol{B} = \{B_{ij}\}$ are denoted by $\boldsymbol{A} \cdot \boldsymbol{B} = \{A_{ij}B_{jk}\}$ and $\boldsymbol{A} : \boldsymbol{B} = A_{ij}B_{ji}$, respectively; $\boldsymbol{a} \otimes \boldsymbol{b} = \{a_i b_j\}$ stands for a dyadic product of vectors $\boldsymbol{a} = \{a_i\}$ and $\boldsymbol{b} = \{b_j\}$. The norm of vector $\boldsymbol{a}$ is designated as $|\boldsymbol{a}| = \sqrt{a_i a_i}$ ; $\boldsymbol{0}$ and $\boldsymbol{I}$ are second-order null and unit tensors; and $\boldsymbol{A}^T$, $\boldsymbol{A}^{-1}$, $\det \boldsymbol{A}$, and $\dot{\boldsymbol{A}}$ are the transpose, inverse, determinant, and material time derivatives of $\boldsymbol{A}$, respectively. $\nabla_{\circ}$ and $\nabla$ are the gradients with respect to the undeformed and deformed configurations, respectively; $\nabla_\circ^2 = \nabla_\circ \cdot \nabla_\circ$ is the Laplacian operation in the undeformed configuration; and := stands for equality by definition.

## 2. Thermodynamic treatment

In particle kinematics, the path line of each particle in a continuous media is specifically described by the vector *r*. Each material point is in the undeformed configuration (*r*=*r*$_0$) at the reference time (*t*=*t*$_0$) and in the deformed configuration *r* at the current time *t*, i.e., *r*=*r*(*r*$_0$,*t*). The motion of the material point is described by the deformation gradient tensor $\boldsymbol{F} = \nabla_\circ \boldsymbol{r} = \boldsymbol{I} + \nabla_\circ \boldsymbol{u}$, where *u*=*r*−*r*$_0$ is the displacement vector.

The thermodynamic laws are presented below for an arbitrary volume $V_0$, which is cut from an actual body, with external surfaces $A_0$ that include cracks. Cracks do not refer to discontinuities in the displacement field but regions with a sharp variation of the order parameter. The global form of the first law of thermodynamics is presented as:

$$\int_{A_0} (\boldsymbol{p}_0 \cdot \boldsymbol{v} - \boldsymbol{h}_0 \cdot \boldsymbol{n}_0) dA_0 + \int_{A_0} \boldsymbol{G}_0 \dot{\phi} \cdot \boldsymbol{n}_0 dA_0 + \int_{V_0} \rho_0 (\boldsymbol{f} \cdot \boldsymbol{v} + r) dV_0 = \frac{d}{dt} \int_{V_0} \rho_0 (U + 0.5 \boldsymbol{v} \cdot \boldsymbol{v}) dV_0 \ . \quad (1)$$

Here $\boldsymbol{p}_0 = \boldsymbol{P} \cdot \boldsymbol{n}_0$ is the traction vector acting on the undeformed area; *P* is the first nonsymmetric Piola-Kirchhoff stress tensor, which is defined based on the undeformed configuration; and $\boldsymbol{n}_0$ is the unit outward normal to the undeformed surface. $\boldsymbol{v} = \dot{\boldsymbol{u}}$ is the material velocity and $\boldsymbol{h}_0$ is the heat flux per unit undeformed area. *U*, *f*, and *r* are internal energy, body force vector and the heat supply, respectively, all per unit mass. The generalized force $\boldsymbol{G}_0 \cdot \boldsymbol{n}_0$ is introduced at the undeformed surface, whose conjugate to produce work is the rate of change of the order parameter $\dot{\phi}$. Without $\boldsymbol{G}_0$, the terms which appear due to the dependence of the thermodynamic free energy on the gradient of the order parameter $\nabla_\circ \phi$ are not balanced for an arbitrary volume. The rate of the total entropy production $S_t$ presents the second law of thermodynamics as a result of combining the Clausius-Duhem inequality and the global entropy balance for the entire volume $V_0$:

$$S_t := \frac{d}{dt} \int_{V_0} \rho_0 s dV_0 - \int_{V_0} \rho_0 \frac{r}{\theta} dV_0 + \int_{A_0} \frac{\boldsymbol{h}_0}{\theta} \cdot \boldsymbol{n}_0 dA_0 \geq 0 \ , \quad (2)$$

where *s* is the entropy per unit mass, and $\theta > 0$ is temperature. The Gauss theorem is utilized to transform the surface integrals into the integrals over volume and, after some simplification, the first and second laws of thermodynamics are obtained as:

$$\int_{V_0} \left( \boldsymbol{P} : \dot{\boldsymbol{F}}^T - \rho_0 \dot{U} - \nabla_\circ \cdot \boldsymbol{h}_0 + \rho_0 r + \nabla_\circ \cdot (\boldsymbol{G}_0 \dot{\phi}) + (\nabla_\circ \cdot \boldsymbol{P} + \rho_0 \boldsymbol{f} - \rho_0 \dot{\boldsymbol{v}}) \cdot \boldsymbol{v} \right) dV_0 = 0 ; \quad (3)$$

$$S_t = \int_{V_0} \left( \rho_0 \dot{s} - \rho_0 \frac{r}{\theta} dV_0 + \nabla_\circ \cdot \frac{\boldsymbol{h}_0}{\theta} \right) dV_0 \geq 0 . \quad (4)$$

According to the principle of material frame-indifference, Eq. (3) should be satisfied independent of the velocity of the observer $v_0$ with respect to a fixed frame. Thus, replacing the velocity $v$ with $v-v_0$ should not affect the energy balance. This results in $\nabla_\circ \cdot \boldsymbol{P} + \rho_0 \boldsymbol{f} - \rho_0 \dot{\boldsymbol{v}} = \boldsymbol{0}$ as the equation of motion. We see that the generalized surface force does not change the equation of motion. Shrinking the arbitrary volume to the infinitesimal volume transforms the global Eq. (3) and Eq. (4) to their local forms:

$$\boldsymbol{P}:\dot{\boldsymbol{F}}^T - \rho_0 \dot{U} - \nabla_\circ \cdot \boldsymbol{h}_0 + \rho_0 r + \nabla_\circ \cdot (\boldsymbol{G}_0 \dot{\phi}) = 0 ; \tag{5}$$

$$\rho_0 \tilde{S}_t := \rho_0 \dot{s} - \rho_0 \frac{r}{\theta} + \nabla_\circ \cdot \frac{\boldsymbol{h}_0}{\theta} = \rho_0 \dot{s} - \rho_0 \frac{r}{\theta} + \frac{1}{\theta} \nabla_\circ \cdot \boldsymbol{h}_0 - \frac{1}{\theta^2} \nabla_\circ \theta \cdot \boldsymbol{h}_0 \geq 0. \tag{6}$$

$\tilde{S}_t$ is the rate of entropy production per unit mass. The local energy dissipation rate per unit mass is defined as:

$$\rho_0 \tilde{D} := \rho_0 \theta \tilde{S}_t = \boldsymbol{P}:\dot{\boldsymbol{F}}^T - \rho_0 \dot{U} + \rho_0 \theta \dot{s} + \nabla_\circ \cdot (\boldsymbol{G}_0 \dot{\phi}) - \frac{1}{\theta} \nabla_\circ \theta \cdot \boldsymbol{h}_0 \geq 0 , \tag{7}$$

where we used Eq. (5) to resolve $-\nabla_\circ \cdot \boldsymbol{h}_0 + \rho_0 r$. Splitting Eq. (7) into the mechanical and thermal parts leads to two more strong inequalities. One is Fourier's inequality $-\frac{1}{\theta} \nabla_\circ \theta \cdot \boldsymbol{h}_0 \geq 0$, and the other is the classical mechanical dissipation inequality with a new term at the end:

$$\rho_0 D := \boldsymbol{P}:\dot{\boldsymbol{F}}^T - \rho_0 \dot{U} + \rho_0 \theta \dot{s} + \nabla_\circ \cdot (\boldsymbol{G}_0 \dot{\phi}) \geq 0. \tag{8}$$

Transforming $U = U(\boldsymbol{F}, s, \phi, \nabla_\circ \phi)$ to $\psi = U - \theta s = \psi(\boldsymbol{F}, \theta, \phi, \nabla_\circ \phi)$ leads to a more convenient form of (mechanical) dissipation inequality to manipulate with:

$$\rho_0 D = \boldsymbol{P}:\dot{\boldsymbol{F}}^T - \rho_0 \dot{\psi} - \rho_0 s \dot{\theta} + \nabla_\circ \cdot (\boldsymbol{G}_0 \dot{\phi}) \geq 0. \tag{9}$$

The last term is evaluated as

$$\nabla_\circ \cdot (\boldsymbol{G}_0 \dot{\phi}) = (\nabla_\circ \cdot \boldsymbol{G}_0) \dot{\phi} + \boldsymbol{G}_0 \cdot \nabla_\circ \dot{\phi} = (\nabla_\circ \cdot \boldsymbol{G}_0) \dot{\phi} + \boldsymbol{G}_0 \cdot (\nabla_\circ \dot{\phi}). \tag{10}$$

Substituting Eq. (10) into Eq. (9) and differentiating $\psi$ with respect to all of its variables give:

$$\rho_0 D = \left(\boldsymbol{P} - \rho_0 \frac{\partial \psi}{\partial \boldsymbol{F}}\right):\dot{\boldsymbol{F}}^T - \rho_0 \left(s + \frac{\partial \psi}{\partial \theta}\right)\dot{\theta} - \left(\rho_0 \frac{\partial \psi}{\partial \phi} - \nabla_\circ \cdot \boldsymbol{G}_0\right)\dot{\phi} + \left(\boldsymbol{G}_0 - \rho_0 \frac{\partial \psi}{\partial \nabla_\circ \phi}\right) \cdot (\nabla_\circ \dot{\phi}) \geq 0. \tag{11}$$

By assuming that the dissipation rate depends only on $\dot{\phi}$, entropy $s = -\partial \psi / \partial \theta$; Eq. (12) as an explicit expression for the generalized force; and Eq. (13) for the constitutive equations for the stress tensor, are obtained:

$$\boldsymbol{G}_0 = \rho_0 \frac{\partial \psi}{\partial \boldsymbol{\nabla}_{\!\!\circ} \phi}; \tag{12}$$

$$\boldsymbol{P} = \rho_0 \frac{\partial \psi}{\partial \boldsymbol{F}}. \tag{13}$$

It can be assumed that the dissipation rate also depends on $(\boldsymbol{\nabla}_{\!\!\circ} \dot{\phi})$, and a dissipative contribution to the generalized force $\boldsymbol{G}_0$ may be added. Then, a traditional structure of the Ginzburg-Landau equation will not be obtained. Dissipative stresses such as viscosity can then be added in Eq. (13), as it was done for phase transformations in Levitas (2013b, 2014). However, we would like to focus on fracture as the only dissipation mechanism. Therefore, the only residual term in Eq. (11) is a product of the dissipative force per unit mass $X$ conjugated to $\dot{\phi}$ as follows:

$$\rho_0 D = \rho_0 X \dot{\phi} \geq 0; \quad X := -\frac{\partial \psi}{\partial \phi} + \frac{1}{\rho_0} \boldsymbol{\nabla}_{\!\!\circ} \cdot \left( \rho_0 \frac{\partial \psi}{\partial \boldsymbol{\nabla}_{\!\!\circ} \phi} \right). \tag{14}$$

Eq. (14)$_2$ is the driving force for the evolution of the order parameter $\phi$.

### 3. Boundary condition

The generalized force at the external surface is assumed to be zero as a boundary condition, which is similar to the isolated boundary in the heat conduction problem:

$$\boldsymbol{n}_0 \cdot \boldsymbol{G}_0 = \boldsymbol{n}_0 \cdot \rho_0 \frac{\partial \psi}{\partial \boldsymbol{\nabla}_{\!\!\circ} \phi} = 0. \tag{15}$$

Using Nanson's equation $dA\boldsymbol{n} = dA_0 J \boldsymbol{F}^{-T} \cdot \boldsymbol{n}_0$ (Lai, Rubin, & Krempl, 2009), where the Jacobian $J$ is defined traditionally as $J := dV/dV_0 = \rho_0/\rho = \det \boldsymbol{F}$, the boundary condition in Eq. (15) can be expressed in the deformed configuration:

$$\begin{aligned}\boldsymbol{n}_0 \cdot \boldsymbol{G}_0 dA_0 &= \boldsymbol{n}_0 \cdot \rho_0 \frac{\partial \psi}{\partial \boldsymbol{\nabla}_{\!\!\circ} \phi} dA_0 = J^{-1} \boldsymbol{n} \cdot \boldsymbol{F} \cdot \rho_0 \frac{\partial \psi}{\partial \boldsymbol{\nabla}_{\!\!\circ} \phi} dA = \\ \boldsymbol{n} \cdot \boldsymbol{F} \cdot \rho \frac{\partial \psi}{\partial \boldsymbol{\nabla}_{\!\!\circ} \phi} dA &= \boldsymbol{n} \cdot \boldsymbol{F} \cdot \boldsymbol{F}^{-1} \cdot \rho \frac{\partial \psi}{\partial \boldsymbol{\nabla} \phi} dA = \boldsymbol{n} \cdot \rho \frac{\partial \psi}{\partial \boldsymbol{\nabla} \phi} dA = \boldsymbol{n} \cdot \boldsymbol{G} dA,\end{aligned} \tag{16}$$

where $\boldsymbol{G} := \rho \partial \psi / \partial \boldsymbol{\nabla} \phi$ is the generalized force conjugated to the order parameter at the surface in the deformed configuration and $\partial \psi / \partial \boldsymbol{\nabla}_{\!\!\circ} \phi = \boldsymbol{F}^{-1} \cdot \partial \psi / \partial \boldsymbol{\nabla} \phi$ was used for the last transformation. Finally, the boundary condition in the deformed configuration has the same form as that in the undeformed configuration:

$$\boldsymbol{n} \cdot \boldsymbol{G} = \boldsymbol{n} \cdot \rho \frac{\partial \psi}{\partial \boldsymbol{\nabla} \phi} = 0. \tag{17}$$

Thus, the normal component of the generalized force in the deformed configuration is zero as well.

## 4. Expression of free energy

The Ginzburg-Landau free energy per unit mass is presented in the form of

$$\psi = \psi(\boldsymbol{F}, \phi, \nabla_\circ \phi) = \frac{dS}{dS_0}(\psi^\nabla + \psi^c) + \psi^e; \quad \frac{dS}{dS_0} = J\left|\boldsymbol{F}^{-T} \cdot \boldsymbol{m}_0\right|, \tag{18}$$

where $\psi^\nabla$, $\psi^c$, $\psi^e$ are the gradient, cohesion and elastic parts of the free energy and are all defined per unit mass and calculated in the undeformed volume; $\boldsymbol{m}_0 = \nabla_\circ \phi / |\nabla_\circ \phi|$ is a unit vector normal to the crack surface, i.e., orthogonal to the constant $\phi$ surfaces. $\boldsymbol{k}_0$ and $\boldsymbol{t}_0$ are the mutually-orthogonal unit vectors that are also orthogonal to $\boldsymbol{m}_0$ (see. Fig. 1a); $\boldsymbol{m} = \nabla \phi / |\nabla \phi|$ is a unit vector in the direction of $\nabla \phi$; and $\boldsymbol{k}$ and $\boldsymbol{t}$ are defined in the current configuration, which are the mutually-orthogonal unit vectors both orthogonal to $\boldsymbol{m}$ (see. Fig. 1b).

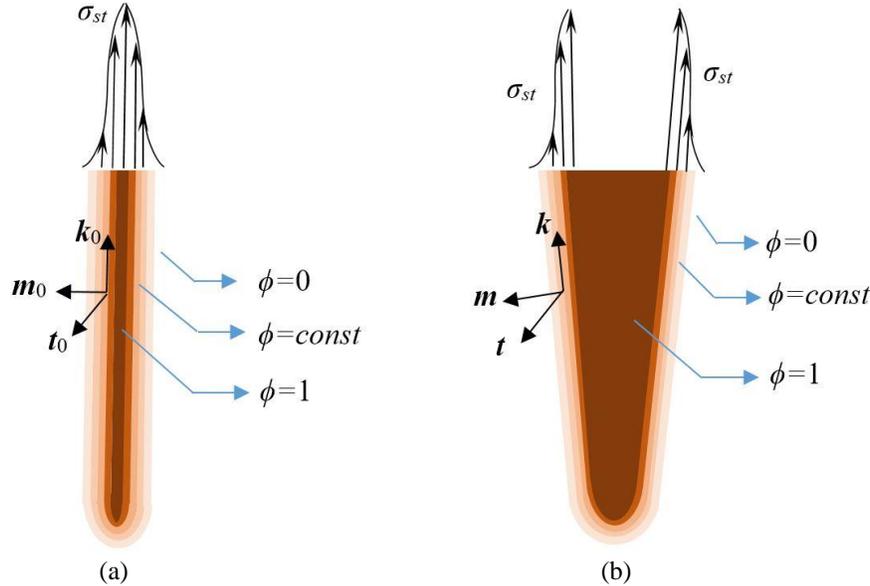

(a)  (b)

**Fig. 1.** a) Schematics of crack with the finite width surfaces described by the level surfaces of the order parameter $\phi =const$ with the distribution of the surface Cauchy stresses in the current configuration. b) Mapping of (a) into the reference configuration. Unit vectors $\boldsymbol{m}$ and $\boldsymbol{m}_0$ are normal to the crack surfaces and mutually orthogonal unit vectors $\boldsymbol{k}$ and $\boldsymbol{t}$, as well as $\boldsymbol{k}_0$ and $\boldsymbol{t}_0$ are within the crack surface. Surface Cauchy stresses are zero along $\boldsymbol{m}$ and have the same components $\sigma^{st}$ along $\boldsymbol{k}$ and $\boldsymbol{t}$.

The multiplier $dS/dS_0 = J\left|\boldsymbol{F}^{-T} \cdot \boldsymbol{m}_0\right|$, which is the ratio of the current to the initial elemental area at the crack surface, is included to obtain biaxial surface stresses with the magnitude of the resultant force equal to the surface energy (see Fig. 1 and Eqs. (27)-(31) below). The gradient and the cohesive energies are both localized at the diffuse crack surfaces, and their sum determines the surface energy. The energy of the elemental volume at one crack surface (defined as the excess energy with respect to the bulk material, without elastic energy) with a spatial coordinate $\xi_0$ along $\boldsymbol{m}_0$ is

$$d\Gamma = 0.5 \int_{-\infty}^{+\infty} \frac{dS}{dS_0}(\psi^\nabla + \psi^c)\rho_0 d\xi_0 dS_0 = 0.5 \left( \int_{-\infty}^{+\infty} \rho_0(\psi^\nabla + \psi^c) d\xi_0 \right) dS = \gamma dS;$$

$$\gamma = 0.5 \int_{-\infty}^{+\infty} \rho_0(\psi^\nabla + \psi^c) d\xi_0 = \rho_0(\psi^\nabla + \psi^c)_{av} h_0,$$

(19)

where the crack's possession of two material surfaces is taken into account, $h_0$ is the width of the crack surface (in which $\phi > 0$) in the reference configuration, and the subscript "*av*" denotes averaging over one crack surface. Eq. (19) shows that the term $\gamma$ is the surface energy per unit deformed area. Without the multiplier $dS/dS_0$, this term would define the surface energy per unit undeformed area $\gamma_0$. Thus, instead of $\gamma_0 dS_0$, the term $\gamma dS$ is introduced, producing surface tension in classical thermodynamics (see, e.g., (Porter et al., 2009)). Although $dS/dS_0$ in the small strain theory is close to unity, it provides a finite contribution to the derivative of the free energy with respect to strains, i.e., to stresses. Therefore, even in the small strain theory, some of the geometric nonlinearities are retained for the reproduction of the surface stresses (Levitas et al., 2018). The expression of each energy term is described below.

*Gradient energy* is defined in the undeformed configuration and is accepted in the conventional form:

$$\rho_0 \psi^\nabla = \rho_0 \psi^\nabla (\nabla_\circ \phi) = \frac{1}{2} \beta (\nabla_\circ \phi)^2.$$  (20)

*Cohesion energy* is expressed as

$$\rho_0 \psi^c = \rho_0 \psi^c(\phi) = A f(\phi),$$  (21)

where $A$ is the maximum cohesion energy corresponding to the fully broken bonds; $f(\phi)$ is an interpolation function for the cohesion energy and will be determined below.

*Elastic energy* is expressed as the Taylor series of the elastic Lagrangian strain tensor $\boldsymbol{E} = (\boldsymbol{F}^T \cdot \boldsymbol{F} - \boldsymbol{I})/2$. The $k^{th}$-rank elastic moduli tensors $\boldsymbol{C}_k$ are degraded by the degradation function, $I(\phi)$:

$$\rho_0 \psi^e = \rho_0 \psi^e(\phi, \boldsymbol{F}) := \rho_0 \bar{\psi}^e(\phi, \boldsymbol{E}) =$$
$$I(\phi) \left( \frac{1}{2} \boldsymbol{E} : \boldsymbol{C}_2 : \boldsymbol{E} + \frac{1}{3!}(\boldsymbol{E} : \boldsymbol{C}_3 : \boldsymbol{E}) : \boldsymbol{E} + \frac{1}{4!} \boldsymbol{E} : (\boldsymbol{E} : \boldsymbol{C}_4 : \boldsymbol{E}) : \boldsymbol{E} + ... \right) := I(\phi) \Psi^e(\boldsymbol{E}),$$  (22)

where $\Psi^e$ is the elastic energy of the damage-free material. In general, each $\boldsymbol{C}_k$ can have a different degradation function.

## 5. Expression of stress tensors

Eqs. (13) is used to obtain the first Piola-Kirchhoff $\boldsymbol{P}$ and Cauchy $\boldsymbol{\sigma} = \boldsymbol{P} \cdot \boldsymbol{F}^T / J$ stress tensors. Stress tensors are split into an elastic part (with superscript *e*) and a surface part (with superscript *st*):

$$\boldsymbol{P} = \boldsymbol{P}^e + \boldsymbol{P}^{st}; \qquad \boldsymbol{\sigma} = \boldsymbol{\sigma}^e + \boldsymbol{\sigma}^{st}. \tag{23}$$

*Elastic stresses.* The elastic part of the stress tensor is obtained according to the conventional definition:

$$\boldsymbol{P}^e = \rho_0 \frac{\partial \psi^e}{\partial \boldsymbol{F}} = \rho_0 \boldsymbol{F} \cdot \frac{\partial \psi^e}{\partial \boldsymbol{E}}; \tag{24}$$

with

$$\frac{\partial \psi^e}{\partial \boldsymbol{E}} = I(\phi)\left(\boldsymbol{C}_2 : \boldsymbol{E} + \frac{1}{2}\boldsymbol{E} : \boldsymbol{C}_3 : \boldsymbol{E} + \frac{1}{3!}(\boldsymbol{E} : \boldsymbol{C}_4 : \boldsymbol{E}) : \boldsymbol{E} + ...\right), \tag{25}$$

where $\partial \psi^e / \partial \boldsymbol{F} = \boldsymbol{F} \cdot \partial \psi^e / \partial \boldsymbol{E}$ is used. The Cauchy elastic stress is:

$$\boldsymbol{\sigma}_e = \frac{1}{J}\boldsymbol{P}^e \cdot \boldsymbol{F}^T = \frac{\rho_0}{J}\boldsymbol{F} \cdot \frac{\partial \psi^e}{\partial \boldsymbol{E}} \cdot \boldsymbol{F}^T = \rho \boldsymbol{F} \cdot \frac{\partial \psi^e}{\partial \boldsymbol{E}} \cdot \boldsymbol{F}^T. \tag{26}$$

*Surface stresses.* Since $dS/dS_0$ depends on $\boldsymbol{F}$, it produces a finite contribution to the stresses. This leads to a desirable expression for the surface stresses. Thus,

$$\boldsymbol{P}^{st} = \rho_0 (\psi^{\nabla} + \psi^c)\frac{d}{d\boldsymbol{F}}\frac{dS}{dS_0} = \rho_0 \frac{dS}{dS_0}(\psi^{\nabla} + \psi^c)(\boldsymbol{I} - \boldsymbol{m} \otimes \boldsymbol{m}) \cdot \boldsymbol{F}^{-T}. \tag{27}$$

Here $\boldsymbol{m}$ is normal to the constant $\phi$ surfaces, i.e., crack surface (see Fig. 1b) and the detailed derivation is shown in the Appendix (Eqs. (108)-(113)).

The true surface stresses are obtained as

$$\boldsymbol{\sigma}^{st} = \frac{1}{J}\boldsymbol{P}^{st} \cdot \boldsymbol{F}^T = \frac{\rho_0}{J}\frac{dS}{dS_0}(\psi^{\nabla} + \psi^c)(\boldsymbol{I} - \boldsymbol{m} \otimes \boldsymbol{m}) = \rho \frac{dS}{dS_0}(\psi^{\nabla} + \psi^c)(\boldsymbol{I} - \boldsymbol{m} \otimes \boldsymbol{m}). \tag{28}$$

They represent an isotropic biaxial tension along the crack surface with the magnitude of

$$\sigma^{st} := \rho \frac{dS}{dS_0}(\psi^{\nabla} + \psi^c). \tag{29}$$

Thus, $dS/dS_0$ is multiplied to those terms of the free energy, which we would like to contribute to the biaxial part of the Cauchy stress tensor. This is more evident in the small strain framework, when $dS/dS_0 = 1 + (\boldsymbol{I} - \boldsymbol{m} \otimes \boldsymbol{m}):\boldsymbol{\varepsilon}$, and $d(dS/dS_0)/d\boldsymbol{\varepsilon} = \boldsymbol{I} - \boldsymbol{m} \otimes \boldsymbol{m}$, where $\boldsymbol{\varepsilon}$ is the small strain tensor (Levitas et al., 2018).

Operating with parameters averaged over the undeformed crack surface width $h_0$, we obtain:

$$\sigma^{st}_{av}dV = \sigma^{st}_{av}\frac{\rho_0}{\rho_{av}}dV_0 = \sigma^{st}_{av}\frac{\rho_0}{\rho_{av}}h_0 dS_0 = \rho_{av}\frac{dS}{dS_0}(\psi^{\nabla} + \psi^c)_{av}\frac{\rho_0}{\rho_{av}}h_0 dS_0 =$$
$$\rho_0(\psi^{\nabla} + \psi^c)_{av}h_0 dS = \gamma dS, \tag{30}$$

i.e., $\sigma_{av}^{st} h = \gamma$ where $h := dV \div dS$ is the width of the crack surface in the actual configuration. The resultant force acting at each crack surface $T$ with a spatial coordinate $\xi$ along $\boldsymbol{m}$ is

$$T = 0.5 \int_{-\infty}^{+\infty} \sigma_{st} d\xi = 0.5 \int_{-\infty}^{+\infty} \rho \frac{dS}{dS_0} (\psi^\nabla + \psi^c) d\xi = 0.5 \int_{-\infty}^{+\infty} (\psi^\nabla + \psi^c) \frac{dm}{dS_0} = $$
$$0.5 \int_{-\infty}^{+\infty} \rho_0 (\psi^\nabla + \psi^c) d\xi_0 = \gamma, \qquad (31)$$

where $dm = \rho dS d\xi = \rho_0 dS_0 d\xi_0$ is the elemental mass. Thus, introducing the factor $dS/dS_0$ in the expression for energy, we obtained the isotropic biaxial surface tension with the resultant force equal to the surface energy per unit crack surface in the actual configuration.

*Total stresses.* Combining all the contributions, we obtain the total first Piola-Kirchhoff stress tensor as

$$\boldsymbol{P} = \boldsymbol{P}^e + \boldsymbol{P}^{st} = \rho_0 \boldsymbol{F} \cdot \frac{\partial \psi^e}{\partial \boldsymbol{E}} + \rho_0 \frac{dS}{dS_0} (\psi^\nabla + \psi^c)(\boldsymbol{I} - \boldsymbol{m} \otimes \boldsymbol{m}) \cdot \boldsymbol{F}^{-T}, \qquad (32)$$

where we used the total true stress tensor as

$$\boldsymbol{\sigma} = \boldsymbol{\sigma}^e + \boldsymbol{\sigma}^{st} = \rho \boldsymbol{F} \cdot \frac{\partial \psi^e}{\partial \boldsymbol{E}} \cdot \boldsymbol{F}^T + \rho \frac{dS}{dS_0} (\psi^\nabla + \psi^c)(\boldsymbol{I} - \boldsymbol{m} \otimes \boldsymbol{m}). \qquad (33)$$

Stresses $\boldsymbol{P}_{st}$ and $\boldsymbol{\sigma}_{st}$ are called the structural stresses at the surface because $\boldsymbol{\sigma}_{st}$ reduces to a biaxial stress tensor with the magnitude of the resultant force equal to the surface energy per unit current area (see Eq. (31)).

*Surface stresses in small strain theory.* In the limit of small strains and rotations, one has $dS/dS_0 \simeq 1$ as well as $\rho \simeq \rho_0$, while Eq. (27) and Eq. (28)$_1$ reduce to:

$$\boldsymbol{\sigma}^{st} = \boldsymbol{P}^{st} = \rho (\psi^\nabla + \psi^c)(\boldsymbol{I} - \boldsymbol{m} \otimes \boldsymbol{m}) \neq \boldsymbol{0}, \qquad (34)$$

which are equal to the surface stresses in Levitas et al. (2018). Thus, while the large strain formulation was required to introduce the surface stresses, the surface stresses do not disappear or even change at small strains. This is why incorporating the large strain formulation is essential, even for a small strain study, to introduce the surface stresses.

### 6. Ginzburg-Landau equation

The time-dependent Ginzburg-Landau equation has the same origin for and application to any structural changes: it is a linear relationship between the order parameter rate and the thermodynamically-conjugate thermodynamic force. The linear relationship between $X$ and $\dot{\phi}$ leads to the generalized Ginzburg-Landau equation accounting for the surface stresses.

$$\dot{\phi}(\boldsymbol{r}_0, t) = L \left[ -\frac{\partial \psi}{\partial \phi} + \nabla_\circ \cdot \frac{\partial \left( (dS/dS_0)(\psi^\nabla + \psi^c) \right)}{\partial \nabla_\circ \phi} \right], \qquad (35)$$

where $L$ is the kinetic coefficient, this guarantees that Eq. (14)$_1$ is always satisfied. Note that the initial homogeneity in density where $\nabla_\circ \rho_0 = \boldsymbol{0}$ was assumed in writing Eq. (35).

### 6.1. Undeformed configuration

By elaborating Eq. (35), an explicit evolution for the order parameter is obtained; this is a generalization of the Ginzburg-Landau to include the surface effects.

$$\frac{1}{\bar{L}}\dot{\phi}(\boldsymbol{r}_0,t) + I'(\phi)\Psi^e + A\frac{dS}{dS_0}f'(\phi) = \nabla_\circ \cdot \frac{\partial\big((dS/dS_0)(\psi^\nabla + \psi^c)\big)}{\partial \nabla_\circ \phi} =$$

$$\nabla_\circ \cdot \left(\frac{dS}{dS_0}\frac{\partial\big(0.5\beta(\nabla_\circ \phi)^2 + \psi^c\big)}{\partial \nabla_\circ \phi} + \frac{\partial(J|\boldsymbol{F}^{-T}\cdot \boldsymbol{m}_0|)}{\partial \nabla_\circ \phi}(\psi^\nabla + \psi^c)\right) =$$

$$\nabla_\circ \cdot \big(\beta(dS/dS_0)\nabla_\circ \phi + J\boldsymbol{M}(\psi^\nabla + \psi^c)\big) = \beta(dS/dS_0)\nabla_\circ^2\phi + \beta\nabla_\circ\phi\cdot\nabla_\circ(dS/dS_0) + \tag{36}$$

$$(J\nabla_\circ\cdot\boldsymbol{M} + \boldsymbol{M}\cdot\nabla_\circ J)(\psi^\nabla + \psi^c) + J\boldsymbol{M}\cdot\nabla_\circ\big(0.5\beta(\nabla_\circ\phi)^2 + Af(\phi)\big) =$$

$$\beta(dS/dS_0)\nabla_\circ^2\phi + \beta\nabla_\circ\phi\cdot\boldsymbol{Y} + J\big(N + \boldsymbol{M}\cdot(\boldsymbol{F}^{-T}:\nabla_\circ\boldsymbol{F})\big)(\psi^\nabla + \psi^c) +$$

$$\beta J\boldsymbol{M}\cdot\nabla_\circ\nabla_\circ\phi\cdot\nabla_\circ\phi + JAf'(\phi)\boldsymbol{M}\cdot\nabla_\circ\phi,$$

where $\bar{L} := L/\rho_0$ and $\nabla_\circ J = dJ/d\boldsymbol{r} = (dJ/d\boldsymbol{F}):d\boldsymbol{F}/d\boldsymbol{r} = J\boldsymbol{F}^{-T}:\nabla_\circ\boldsymbol{F}$ is used. Derivations and final expressions of $\boldsymbol{M}$, $\boldsymbol{Y}$, and $N$ with the below definitions are given in Appendix (Eqs. (114)-(123)).

$$\boldsymbol{M} := \frac{\partial}{\partial \nabla_\circ\phi}\big|\boldsymbol{F}^{-T}\cdot \boldsymbol{m}_0\big|, \quad \boldsymbol{Y} := \nabla_\circ \frac{dS}{dS_0}, \quad N := \nabla_\circ\cdot\boldsymbol{M}. \tag{37}$$

Finally, using Eqs. (114)-(123), the Ginzburg-Landau equation in the undeformed configuration Eq. (36) takes the form of

$$\frac{1}{\bar{L}}\dot{\phi}(\boldsymbol{r}_0,t) + I'(\phi)\Psi^e + A\frac{dS}{dS_0}f'(\phi) = \beta\frac{dS}{dS_0}\nabla_\circ^2\phi +$$

$$J\beta\nabla_\circ\phi\cdot\left(\boldsymbol{m}_0\cdot\nabla_\circ\boldsymbol{F}^{-1}\cdot\frac{\boldsymbol{F}^{-T}\cdot\boldsymbol{m}_0}{|\boldsymbol{F}^{-T}\cdot\boldsymbol{m}_0|} + |\boldsymbol{F}^{-T}\cdot\boldsymbol{m}_0|\boldsymbol{F}^{-T}:\nabla_\circ\boldsymbol{F}\right) + J(\psi^\nabla + \psi^c)\times$$

$$\left\{\big(\nabla_\circ\boldsymbol{m}_0\cdot\boldsymbol{F}^{-1}\cdot\boldsymbol{F}^{-T} + \boldsymbol{m}_0\cdot\nabla_\circ\boldsymbol{F}^{-1}\cdot\boldsymbol{F}^{-T} + \boldsymbol{m}_0\cdot\boldsymbol{F}^{-1}\cdot\nabla_\circ\boldsymbol{F}^{-T}\big):\frac{\boldsymbol{I}-\boldsymbol{m}_0\otimes\boldsymbol{m}_0}{|\boldsymbol{F}^{-T}\cdot\nabla_\circ\phi|} - \right. \tag{38}$$

$$\frac{\nabla_\circ\cdot\boldsymbol{m}_0}{|\boldsymbol{F}^{-T}\cdot\nabla_\circ\phi|}\boldsymbol{m}_0\cdot\boldsymbol{F}^{-1}\cdot\boldsymbol{F}^{-T}\cdot\boldsymbol{m}_0 -$$

$$\left.\frac{\boldsymbol{F}^{-T}\cdot\nabla_\circ\phi}{(\boldsymbol{F}^{-T}\cdot\nabla_\circ\phi)^2}\cdot\big(\nabla_\circ\boldsymbol{F}^{-T}\cdot\nabla_\circ\phi + \boldsymbol{F}^{-T}\cdot\nabla_\circ\nabla_\circ\phi\big)\cdot\big(\boldsymbol{I}-\boldsymbol{m}_0\otimes\boldsymbol{m}_0\big)\cdot\boldsymbol{F}^{-1}\cdot\boldsymbol{F}^{-T}\cdot\boldsymbol{m}_0\right\}+$$

$$Jm_0 \cdot F^{-1} \cdot F^{-T} \cdot \frac{I - m_0 \otimes m_0}{|F^{-T} \cdot \nabla_\circ \phi|} \cdot \left( \left( F^{-T} : \nabla_\circ F \right)(\psi^\nabla + \psi^c) + \beta \nabla_\circ \nabla_\circ \phi \cdot \nabla_\circ \phi \right),$$

As can be seen in Eq. (38), distinguishing between deformed and undeformed surfaces directly affects the driving force of the crack nucleation and propagation. Also, the change in the stress distribution due to the contribution of the surface stresses to the mechanical equilibrium equation is another indirect effect of the coefficient $dS/dS_0$ on the Ginzburg-Landau equation.

### 6.1.1. Some simplifications

a) We approximate for small strains and rotations similar to Levitas et al. (2018), i.e., $F \simeq I + \varepsilon + \omega$, and evaluating $F^{-1} \simeq I - \varepsilon - \omega$ and $F^{-T} \simeq I - \varepsilon + \omega$, and neglect all the products of small tensors. $\omega$ is the small rotation tensor, which is the asymmetric part of the gradient of displacement. Then, Substituting Eqs. (125)-(127) into Eq. (36), we obtain the Ginzburg-Landau equation for a small strain framework including the surface effects, which is also given in Levitas et al. (2018):

$$\frac{1}{L}\dot{\phi}(r_0,t) + I'(\phi)\Psi^e + A\left(1 + (I - m_0 \otimes m_0) : \varepsilon\right) f'(\phi) =$$
$$\beta\left(1 + (I - m_0 \otimes m_0) : \varepsilon\right)\nabla_\circ^2 \phi + \beta \nabla_\circ \phi \cdot \nabla_\circ \varepsilon : (I - m_0 \otimes m_0) -$$
$$\beta(1 + \frac{\psi^c}{\psi^\nabla})\left(|\nabla_\circ \phi|\nabla_\circ m_0 : \left(\varepsilon \cdot (I - m_0 \otimes m_0)\right)\right) + \nabla_\circ \phi \cdot \nabla_\circ \varepsilon : (I - m_0 \otimes m_0) + \quad (39)$$
$$m_0 \cdot \varepsilon \cdot \left(m_0 \nabla_\circ^2 \phi - m_0 \cdot (\nabla_\circ \nabla_\circ \phi)^T\right) - \nabla_\circ \phi \cdot \varepsilon \cdot (I - m_0 \otimes m_0) \cdot (\varepsilon : \nabla_\circ \varepsilon) -$$
$$2\beta m_0 \cdot \varepsilon \cdot (I - m_0 \otimes m_0) \cdot \nabla_\circ \nabla_\circ \phi \cdot m_0,$$

b) By neglecting strains but retaining the gradient of small strains, i.e., for $F \simeq I$ but retaining all the gradients, we obtain

$$\frac{1}{L}\dot{\phi}(r_0,t) + I'(\phi)\Psi^e + Af'(\phi) = \beta \nabla_\circ^2 \phi + \beta \nabla_\circ \phi \cdot \left(m_0 \cdot \nabla_\circ F^{-1} \cdot m_0 + \nabla_\circ \cdot F\right) +$$
$$0.5\beta\left(1 + \frac{\psi^c}{\psi^\nabla}\right) \nabla_\circ \phi \cdot \left(\nabla_\circ F^{-1} + \nabla_\circ F^{-T}\right) : (I - m_0 \otimes m_0), \quad (40)$$

where $I : \nabla_\circ F = \nabla_\circ \cdot F$ is used.

c) Keeping all terms except the gradients of $F$ leads to

$$\frac{1}{L}\dot{\phi}(r_0,t) + I'(\phi)\Psi^e + A\frac{dS}{dS_0}f'(\phi) = \beta \frac{dS}{dS_0}\nabla_\circ^2 \phi + J(\psi^\nabla + \psi^c) \times$$
$$\left\{ \nabla_\circ m_0 \cdot F^{-1} \cdot F^{-T} : \frac{I - m_0 \otimes m_0}{|F^{-T} \cdot \nabla_\circ \phi|} - \frac{\nabla_\circ \cdot m_0}{|F^{-T} \cdot \nabla_\circ \phi|} m_0 \cdot F^{-1} \cdot F^{-T} \cdot m_0 - \right. \quad (41)$$

$$\left. \frac{\boldsymbol{F}^{-T}\cdot\nabla_{\circ}\phi}{\left(\boldsymbol{F}^{-T}\cdot\nabla_{\circ}\phi\right)^{2}}\cdot\boldsymbol{F}^{-T}\cdot\nabla_{\circ}\nabla_{\circ}\phi\cdot\left(\boldsymbol{I}-\boldsymbol{m}_{0}\otimes\boldsymbol{m}_{0}\right)\cdot\boldsymbol{F}^{-1}\cdot\boldsymbol{F}^{-T}\cdot\boldsymbol{m}_{0}\right\}+$$

$$J\beta\boldsymbol{m}_{0}\cdot\boldsymbol{F}^{-1}\cdot\boldsymbol{F}^{-T}\cdot\frac{\boldsymbol{I}-\boldsymbol{m}_{0}\otimes\boldsymbol{m}_{0}}{\left|\boldsymbol{F}^{-T}\cdot\nabla_{\circ}\phi\right|}\cdot\nabla_{\circ}\nabla_{\circ}\phi\cdot\nabla_{\circ}\phi.$$

d) For neglected surface stresses ($dS/dS_0 \simeq 1$), the Ginzburg-Landau equation reduces to the standard form:

$$\frac{1}{L}\dot{\phi}(\boldsymbol{r}_{0},t)+I'(\phi)\Psi^{e}+Af'(\phi)=\beta\nabla_{\circ}^{2}\phi. \tag{42}$$

### 6.2. Deformed configuration

The following transformation is used to express the Ginzburg-Landau equation in the deformed configuration:

$$\nabla_{\circ}\cdot\frac{\partial\psi}{\partial\nabla_{\circ}\phi}=\nabla_{\circ}\cdot\left(\frac{\partial\psi}{\partial\nabla\phi}\frac{\partial\nabla\phi}{\partial\nabla_{\circ}\phi}\right)=\nabla_{\circ}\cdot\left(\boldsymbol{F}^{-1}\cdot\frac{\partial\psi}{\partial\nabla\phi}\right)=\nabla\left(\boldsymbol{F}^{-1}\cdot\frac{\partial\psi}{\partial\nabla\phi}\right):\boldsymbol{F}. \tag{43}$$

Then the Ginzburg-Landau equation in the current configuration for an initially homogenous material is

$$\frac{D\phi(\boldsymbol{r},t)}{Dt}=\frac{\partial\phi(\boldsymbol{r},t)}{\partial t}+\boldsymbol{v}\cdot\nabla\phi=L\left[-\frac{\partial\psi}{\partial\phi}+\nabla\left(\boldsymbol{F}^{-1}\cdot\frac{\partial\left((dS/dS_{0})(\psi^{\nabla}+\psi^{c})\right)}{\partial\nabla\phi}\right):\boldsymbol{F}\right], \tag{44}$$

where

$$\rho_{0}\psi^{\nabla}=J\rho\psi^{\nabla}=\frac{1}{2}\beta(\nabla_{\circ}\phi)^{2}=\frac{1}{2}\beta\left(\boldsymbol{F}^{T}\cdot\nabla\phi\right)^{2}, \tag{45}$$

and

$$\frac{dS}{dS_{0}}=J\left|\boldsymbol{F}^{-T}\cdot\boldsymbol{m}_{0}\right|=J\left|\boldsymbol{F}^{-T}\cdot\nabla_{\circ}\phi\right|/\left|\nabla_{\circ}\phi\right|=J\left|\nabla\phi\right|/\left|\boldsymbol{F}^{T}\cdot\nabla\phi\right|=J/\left|\boldsymbol{F}^{T}\cdot\boldsymbol{m}\right|. \tag{46}$$

In Eq. (44), the material time derivative of $\phi$ in the undeformed configuration is transformed to the corresponding expression in the deformed configuration.

### 7. Specification of the cohesion energy and degradation function

The detailed analysis of the homogeneous solution and the main requirements of the cohesion energy and the degradation function can be found in (Levitas et al., 2018). The well-known requirements, are summarized below, as well as new requirements which we want to impose:

a) The only existing energy in the intact state where $\phi = 0$, is the elastic energy. Thus, $f(0)=0$ and $I(0)=1$.
b) The maximum cohesion energy $A$ is reached at the fully damaged state where $\phi = 1$, then, $f(1)=1$.
c) The fully damaged state $\phi = 1$ cannot sustain any elastic energy, i.e., $I(1)=0$.

Thus, $I(\phi)$ is employed as:

$$I(\phi) = (1-\phi)^n; \quad n \geq 1, \tag{47}$$

which satisfies all the three mandatory requirements above; the parameter $n$ is introduced so that different stress-strain curves can be obtained.

d) The homogeneous ($\psi^\nabla = 0$) and stationary ($\dot{\phi} = 0$) state, leads to the equilibrium form of the Ginzburg-Landau equation

$$\frac{\partial \psi}{\partial \phi} = I'(\phi_e)\Psi_e^e + A \frac{dS}{dS_0} f'(\phi_e) = 0. \tag{48}$$

It should be mentioned that, while calculating $dS/dS_0 = J|\boldsymbol{F}^{-T} \cdot \boldsymbol{m}_0|$ for a homogeneously distributed order parameter, $\nabla_\circ \phi = \boldsymbol{0}$ and $\boldsymbol{m}_0 = \nabla_\circ \phi / |\nabla_\circ \phi|$ is undefined. The indeterminacy in the direction of $\boldsymbol{m}$ and the surface stress tensor can be eliminated numerically by setting $dS/dS_0=0$ and, consequently, zero surface stresses $\boldsymbol{\sigma}_{st}=\boldsymbol{0}$ when $\nabla \phi = \boldsymbol{0}$.

When the homogeneous state is considered, we assume that the decohesion (cleavage) plane is known and $\boldsymbol{m}$ is defined as orthogonal to it. Eq. (48) results in the following damage equilibrium condition

$$\Psi_e^e(\boldsymbol{E}) = -A \frac{dS}{dS_0} \frac{f'(\phi_e)}{I'(\phi_e)}, \tag{49}$$

which determines the equilibrium value of the order parameter $\phi_e$ for a given strain tensor $\boldsymbol{E}$. If the right-hand side of Eq. (49) is finite at $\phi_e = 0$, then the damage is absent below this value:

$$\phi_e = 0 \text{ (intact state)} \qquad \text{for} \qquad \Psi_e^e \leq -A \frac{dS^*}{dS_0} \frac{f'(0)}{I'(0)}, \tag{50}$$

where $dS^*$ is the infinitesimal deformed area when $\phi$ starts to grow, i.e., at $\phi_e = 0$ when $\dot{\phi}_e > 0$. Eq. (50) is the damage initiation condition. A similar condition was fulfilled for gradient damage models in (Pham and Marigo, 2013)) where a clear elastic threshold is present, below which the damage parameter does not evolve. After excluding the order parameter, Eq. (49), along with Eq. (24) and Eq. (25) for the elastic stresses, represent the equilibrium stress-strain relationship.

Because $f' \geq 0$ and $I' \leq 0$, if polynomial $f'(\phi)$ starts with the same degree as $I'(\phi)$ for $\phi \to 0$, the right-hand side of Eq. (50) is finite and there is a critical elastic energy at the start of damage. For the degradation function in Eq. (47), $I'(\phi_e) = n(1-\phi_e)^{n-1}$ and $I'(0) = n$. Thus, to have a nonzero value of strain for the damage initiation, $f'(0)$ should be finite. Therefore, we choose

$$f(\phi) = \frac{k\phi^2 + \phi}{k+1} \; ; k \geq 0. \tag{51}$$

For large $k$ we have

$$\lim_{k \to \infty} f(\phi) = \lim_{k \to \infty} \frac{k\phi^2 + \phi}{k} = \lim_{k \to \infty} \left( \phi^2 + \frac{\phi}{k} \right) = \phi^2. \tag{52}$$

Moreover, for $k$ close to 0 we obtain

$$\lim_{k \to 0} f(\phi) = \lim_{k \to 0} (k\phi^2 + \phi) = \phi. \tag{53}$$

Such an analysis on the degradation function in essential, as a new degradation function was introduced in Wilson, Borden, & Landis (2013) to address undesired consequences of the classic quadratic function. Figure 2 shows $I(\phi)$ and $f(\phi)$ for various $n$ and $k$.

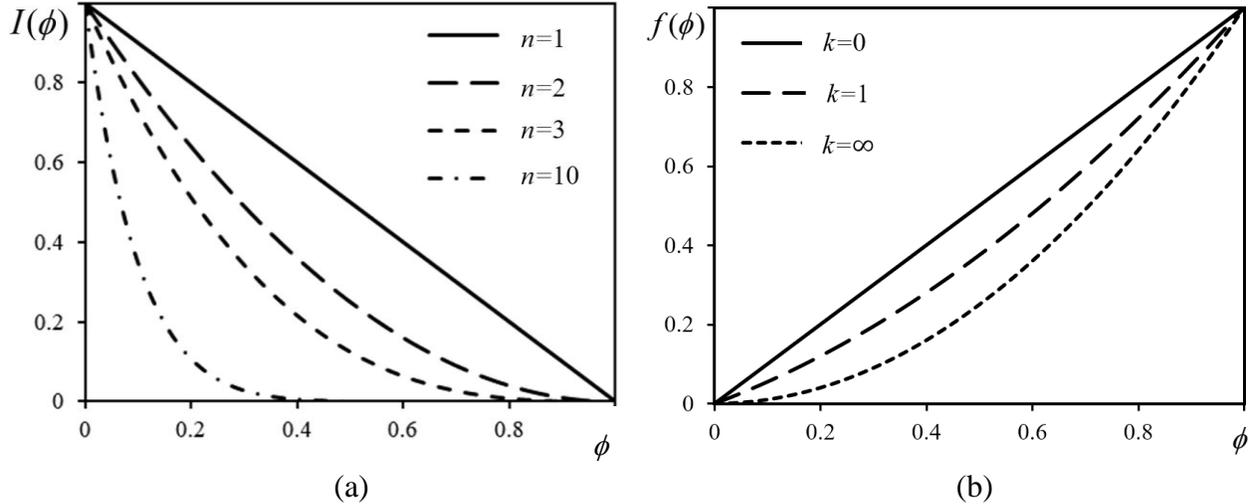

(a) (b)

**Fig. 2.** Effect of the material parameters on the a) degradation function $I(\phi)$ and b) interpolation function $f(\phi)$.

Then Eqs. (49) and (50) are simplified to

$$\Psi_e^e = A \frac{dS}{dS_0} \frac{2k\phi_e + 1}{(k+1)n(1-\phi_e)^{n-1}} \quad \text{for} \quad \Psi_e^e \geq \Psi_i^e;$$

$$\phi_e = 0 \quad \text{for} \quad \Psi_e^e < \Psi_i^e, \tag{54}$$

where, the critical strain (elastic energy), which is the strain (elastic energy) at the initiation of damage (subscript $i$), is found from:

$$\Psi_i^e = \Psi^e(E_i) = A\frac{dS^*}{dS_0}\frac{1}{(k+1)n}. \tag{55}$$

According to Eq. (54) for $\phi_e = 1$, damage completes at finite strain for $n=1$ and at infinite strain for $n>1$.

The stability condition is now checked; because

$$\frac{\partial^2 \psi}{\partial \phi^2} = I''(\phi_e)\Psi_e^e + A\frac{dS}{dS_0}f''(\phi_e) = n(n-1)(1-\phi_e)^{n-2}\Psi_e^e + A\frac{dS}{dS_0}\frac{2k}{k+1} > 0, \tag{56}$$

the equilibrium solution in Eq. (54) is stable and corresponds to the minimum of the free energy during the damage growth.

## 8. Equilibrium stress-strain curves

Here, the relationship between the second Piola-Kirchhoff stress $T$ and the Lagrangian strain $E$ is used in the one-dimensional problem and the simplest quadratic energy, i.e., $\Psi^e = 0.5C_2 E^2$ is considered, without the surface stresses. $C_2$ is Young's modulus. $T$ is then obtained as

$$T = \frac{\partial \psi^e}{\partial E} = I(\phi)C_2 E. \tag{57}$$

Utilizing Eqs. (48) and (57), we obtain

$$T = I\sqrt{-\frac{2C_2 A f'}{I'}}; \tag{58}$$

Now, let us calculate $dT/d\phi$ at $\phi = 0$; thus,

$$\frac{dT}{d\phi} = I'\sqrt{-\frac{2C_2 A f'}{I'}} - I\frac{C_2 A(f''I' - I''f')}{\sqrt{-2C_2 A f I'}}. \tag{59}$$

For $I(\phi) = (1-\phi)^n$ and $f(\phi) = (k\phi^2 + \phi)/(k+1)$, we obtain

$$\frac{dT}{d\phi} = I'\sqrt{-\frac{2C_2 A f'}{I'}} - I\frac{C_2 A\left(I' - I''(2k\phi+1)/(k+1)\right)}{\sqrt{-2C_2 A f I'}} =$$
$$-n(1-\phi)^{n-1}\sqrt{-\frac{2C_2 A f'}{I'}} + I\frac{n(1-\phi)^{n-2}C_2 A\left((1-\phi) - (n-1)(2k\phi+1)/(k+1)\right)}{\sqrt{-2C_2 A f I'}}. \tag{60}$$

For the onset of damage $\phi = 0$:

$$\left.\frac{dT}{d\phi}\right|_{\phi=0} = 0.5\sqrt{nC_2A}\left(1 - \frac{2n}{k+1}\right), \tag{61}$$

which is zero for $2n=k+1$ and negative for $2n>k+1$. For $2n \geq k+1$, since $dE/d\phi > 0$ is always true, the equilibrium tangential modulus $dT/dE \leq 0$ at the onset of the damage. Thus, *damage starts at the strain at which the second Piola-Kirchhoff stress has its maximum, and the tangent modulus jumps from $C_2$ to the non-positive value* (see Fig. 3 for $k=1$ and various $n \geq 1$ and Fig. 6 for $n=1$ and $k \leq 1$). Then, after the initiation of damage and during it, the second Piola-Kirchhoff stress decreases. While not necessarily valid for other stresses and strains (see Fig. 4, 5), this is more realistic than in some previous models (Levitas et al., 2018) in which the damage always starts at an infinitesimal strain and the elastic modulus continuously decreases from its value in the undamaged state.

$T_i$ and $E_i$ can then be calculated by substituting $\Psi^e = 0.5 C_2 E^2$ in Eq. (55), as follows:

$$\Psi_i^e = 0.5 C_2 E_i^2 = A \frac{dS^*}{dS_0} \frac{1}{(k+1)n}, \tag{62}$$

resulting in

$$E_i = \frac{\sqrt{A dS^*/dS_0}}{\sqrt{C_2(k+1)n}}; \qquad T_i = \frac{\sqrt{A C_2 dS^*/dS_0}}{\sqrt{(k+1)n}}, \tag{63}$$

which are equal to the strain and stress at the peak point when $2n \geq k+1$. Since $dS^*/dS_0$ also depends on $E_i$, Eq. (63) is dependent on $E_i$, which can be easily solved for different models. In the first approximation $dS^*/dS_0 \simeq 1$, we obtain explicit relationships

$$E_i = \frac{\sqrt{A}}{\sqrt{C_2(k+1)n}}; \qquad T_i = \frac{\sqrt{A C_2}}{\sqrt{(k+1)n}}. \tag{64}$$

Thus, the strain at the damage initiation and the corresponding stress, in addition to the magnitude of the cohesion energy, can be controlled by the parameters in the degradation function $n$ and in the interpolation function $k$. Eq. (64) is important in the sense that, for given $n$ and $k$, it determines $A$ in terms of the elastic energy (or maximum $E_i$ or $T_i$) at the beginning of damage. As we will see in Eq. (67), $A$ is also the total work until the damage completes. Then, the parameters $n$ and $k$ produce portioning between the works in the elastic region and after the initiation of damage until the fracture completes.

Excluding the order parameter from Eqs. (49) and (57) leads to the equilibrium stress-strain relationship. Figure 3 shows the uniaxial equilibrium second Piola-Kirchhoff stress vs. Lagrangian strain (*T-E*) curves for different values of $n$ and $k=1$.

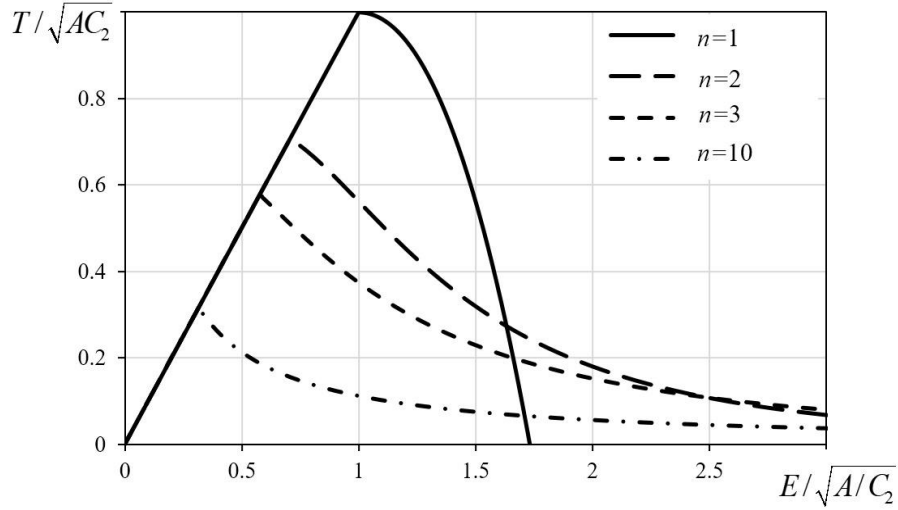

**Fig. 3.** Normalized second Piola-Kirchhoff stress $T$ vs. Lagrangian strain $E$ for uniaxial tension for different $n$ and $k=1$.

Using the relationship between the first Piola-Kirchhoff stress and second Piola-Kirchhoff stress,

$$P = FT = I(\phi)C_2 FE = 0.5I(\phi)C_2 F(F^2 - 1) \;;\qquad F = (1+2E)^{0.5}, \tag{65}$$

we plot the first Piola-Kirchhoff stress against its work-conjugate strain, i.e., $F$, in Fig. 4.

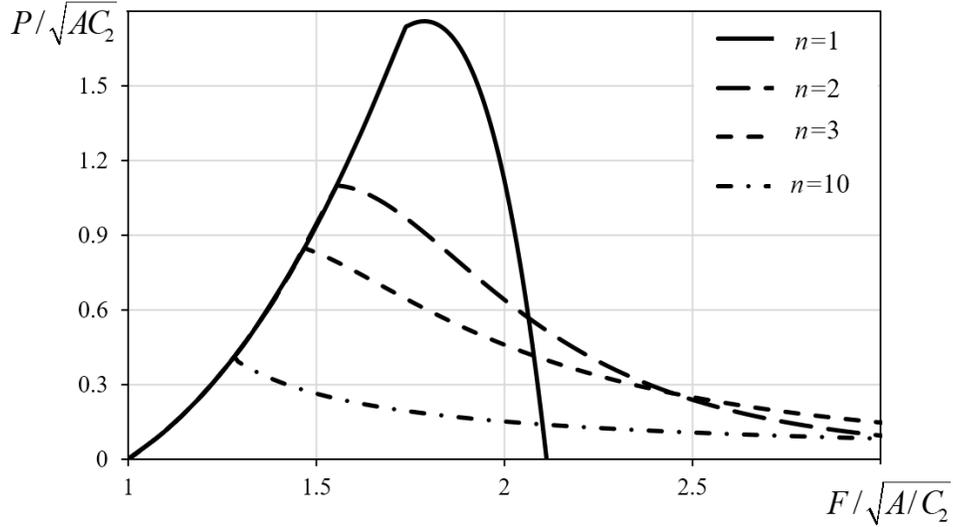

**Fig. 4.** Normalized first Piola-Kirchhoff stress $P$ vs. deformation gradient $F$ for uniaxial tension for different $n$ and $k=1$.

For the uniaxial loading along the normal-to-the-crack direction, the work increment per unit current volume is $J^{-1}PdF = \dfrac{\sigma}{F}dF = \sigma d(\ln F)$, i.e., the work-conjugate of Cauchy stress $\sigma$ is the logarithmic strain $\ln F$. The Cauchy stress $\sigma_1$ in the normal direction is defined as

$$\sigma_1 = J^{-1} P_1 F_1 = P_1 F_1 \frac{dV_0}{dV} = P_1 F_1 \frac{dV_0}{F_1 F_2 F_3 dV_0} = P_1 \frac{1}{F_2 F_3} = P_1 \frac{dS_0}{dS} \quad \text{and} \quad \sigma_1 = F_1 T_1 \frac{dS_0}{dS}. \tag{66}$$

The Cauchy stress $\sigma$ vs. ln $F$ is presented in Fig. 5 under the assumption $dS/dS_0 \simeq 1$. For both $P$ and $\sigma$, the elastic part is nonlinear, and for $n$ close to 1, the stresses continue to slightly grow after the damage starts before they decrease.

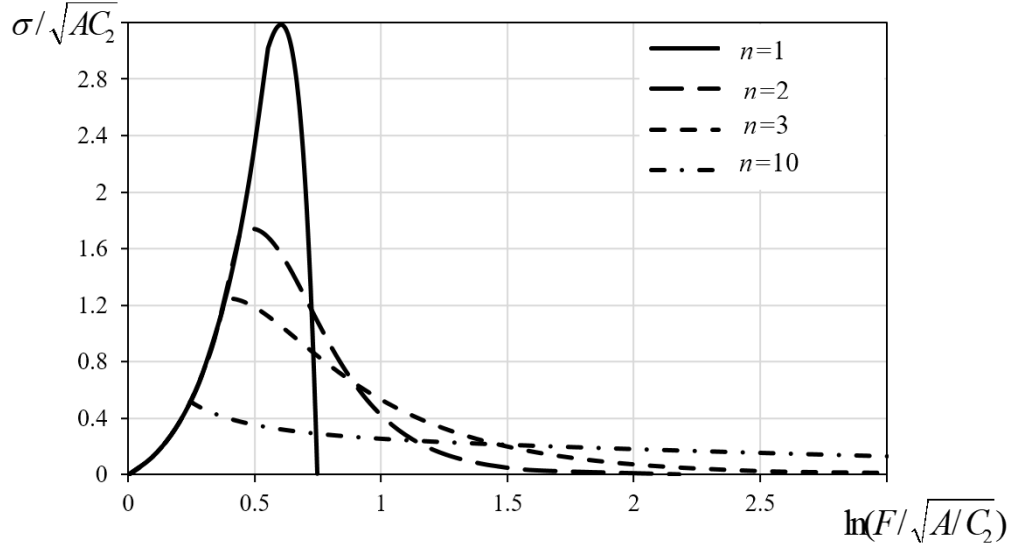

**Fig. 5.** Normalized Cauchy stress $\sigma$ vs. ln $F$ for uniaxial tension for different $n$ and $k=1$.

To investigate the effect of $k$ on stress-strain curves, Figure 6 shows the equilibrium solution of the $T$-$E$ curve for various $k$ and $n=1$.

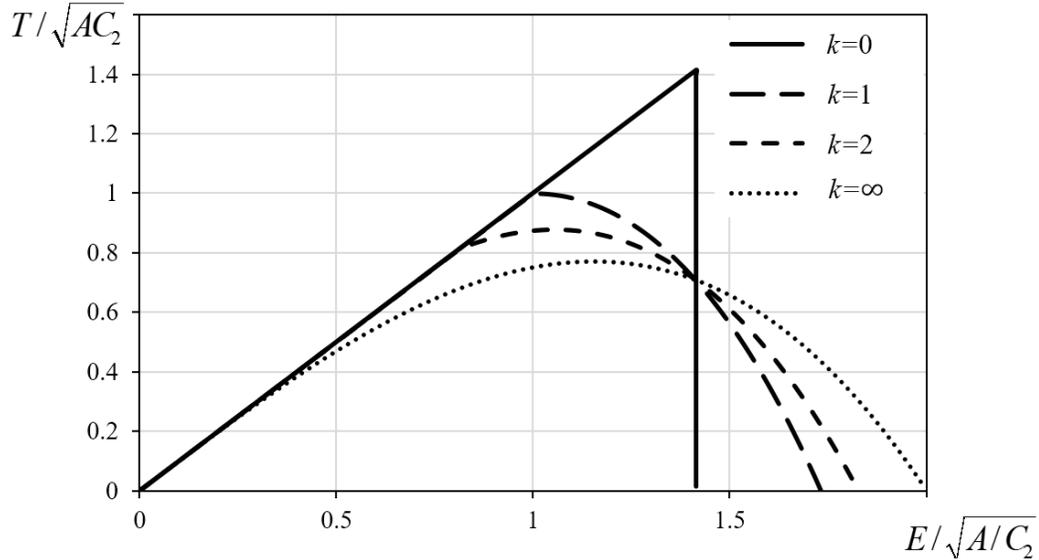

**Fig. 6.** Normalized second Piola-Kirchhoff stress $T$ vs. Lagrangian strain $E$ for uniaxial tension for different $k$ and $n=1$.

As can be deduced from Fig. 6, the peak point and the start of damage are equal and coincide for $k \leq 1$. The general condition for this coincidence was discussed in Eq. (61) and after it. For $k>1$, the damage initiates before the peak point of $T$, producing the nonlinear portion of the $T$-$E$ curve. The damage initiation and the peak point of $T$ both decrease as $k$ increases.

*Calibration of A.* We consider the general three-dimensional homogenous state, and evaluate the elastic work per unit undeformed volume:

$$\int_{F=I}^{F_{\phi=1}} P : dF^T = \int_{F=I}^{F_{\phi=1}} \rho_0 \frac{\partial \psi}{\partial F} : dF^T = \rho_0 \int_{F=I}^{F_{\phi=1}} d\psi(F, \phi) = \rho_0 \left( \psi(F_{\phi=1}, 1) - \psi(I, 0) \right) = \\ I(1)\Psi^e(E_{\phi=1}) - I(0)\Psi^e(\mathbf{0}) + A(S_{\phi=1}/S_0)(f(1) - f(0)) = AS_{\phi=1}/S_0. \tag{67}$$

The equality $d\psi = (\partial \psi / \partial F) : dF^T + (\partial \psi / \partial \phi) d\phi = (\partial \psi / \partial F) : dF^T$ has been used; $\partial \psi / \partial \phi = 0$ is used because of the thermodynamic equilibrium condition; $I(1)=0$; $I(0)=1$; $\Psi^e(\mathbf{0})=0$; $f(0)=0$; and $f(1)=1$. Here $S_0$, and $S_{\phi=1}$ are the surfaces of the crack in the reference configuration and current configuration at $\phi = 1$, respectively. Eq. (67) is valid for any type of nonlinear hyperelastic materials. For one-dimensional homogenous tension, the elastic work is equal to the area under the stress-strain curve.

The elastic work within the reference volume $S_0 d$, where $d$ is the initial thickness of the cohesive layer, should be equal to the created surface energy. Thus, $(AS_{\phi=1}/S_0)S_0 d = 2\gamma S_{\phi=1}$ and the maximum cohesion energy, i.e., the parameter $A$, is obtained as:

$$A = \frac{2\gamma}{d}. \tag{68}$$

Note that, since the normal-to-the-crack surface stress is zero at $\phi = 1$, and if all the other stresses are also zero in the experiment, $S_0 = S_{\phi=1}$ and $\gamma$ is the surface energy per undeformed area.

## 9. Stationary solution

The stationary Ginzburg-Landau equation in the stress-free case is:

$$\beta \nabla_\circ^2 \phi = A \frac{df}{d\phi}, \tag{69}$$

where the deformation of the diffuse surfaces is neglected. Integration of Eq. (69) over $\phi$ (see Levitas et al. (2018) for more details) leads to

$$\frac{\beta}{2}(\nabla_\circ \phi)^2 = Af(\phi) \Rightarrow \psi^\nabla = \psi^c. \tag{70}$$

According to Eq. (70), the excess of the cohesion energy is equal to the gradient energy. Allowing for our specific interpolation function Eq. (51), in the one-dimensional case, Eq. (70) leads to

$$\frac{\beta}{2}\left(\frac{d\phi}{d\xi_0}\right)^2 = \frac{2\gamma}{d} \frac{k\phi^2 + \phi}{k+1}. \tag{71}$$

Finally, the solution of Eq. (71) yields an explicit expression for the two diffuse crack surfaces profile:

$$\begin{cases} \sqrt{k\phi} + \sqrt{k\phi+1} = (\sqrt{k} + \sqrt{k+1})e^{-\sqrt{\frac{k\gamma}{(k+1)\beta d}}|\xi_0|} & |\xi_0| < \xi_{0t}; \\ \phi = 0 & |\xi_0| \geq \xi_{0t}, \end{cases} \quad (72)$$

where $\phi = 1$ corresponds to the separation plane at $\xi_0 = 0$, and

$$\xi_{0t} = \sqrt{\frac{(k+1)\beta d}{k\gamma}} \ln(\sqrt{k} + \sqrt{k+1}) \quad (73)$$

is the transition plane from the damaged state to the intact state. For $k=0$ and infinity, Eqs. (72) and (73) lead to:

$$\begin{cases} \phi = \left(1 - \sqrt{\frac{\gamma}{\beta d}}|\xi_0|\right)^2 & |\xi_0| < \xi_{0t} = \sqrt{\frac{\beta d}{\gamma}}; \\ \phi = 0 & |\xi_0| \geq \xi_{0t}, \end{cases} \quad \text{for } k=0 \text{ and} \quad (74)$$

$$\phi = e^{-2\sqrt{\frac{\gamma}{\beta d}}|\xi_0|}; \qquad \xi_{0t} = \infty, \qquad \text{for } k=\infty.$$

Thus, the intact phase is at a finite distance of $\xi_{0t}$ from the separation plane unless $k$ is infinity. This is in contrast to the other models (Levitas et al., 2018). Figure 7 represents a damage distribution of the crack in the reference configuration, at the position $x_0=0$ in an infinite bar for various $k$.

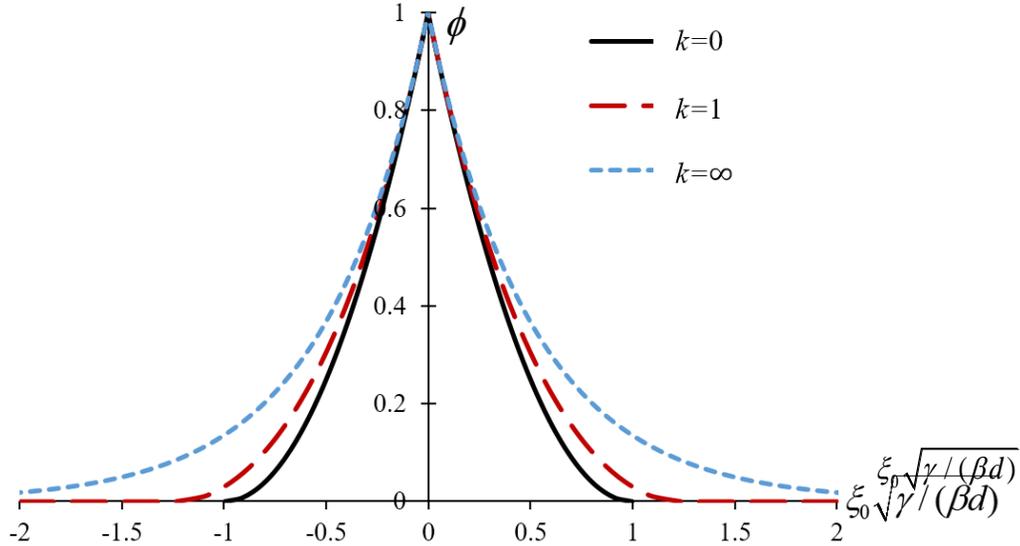

**Fig. 7.** Finite-width profile of the crack surfaces. $\xi_0>0$ is one surface, $\xi_0=0$ is the separation plane, and $\xi_0<0$ is another surface. Damage zone has a finite width $\xi_{0t} = \sqrt{(k+1)\beta d/(k\gamma)} \ln(\sqrt{k} + \sqrt{k+1})$ from each side.

In the general case,

$$\frac{\beta}{2}(\frac{d\phi}{d\xi_0})^2 = \frac{2\gamma}{d}f(\phi) \Rightarrow \int_1^\phi \frac{d\phi}{\sqrt{f(\phi)}} = -2\sqrt{\frac{\gamma}{d\beta}}|\xi_0|. \tag{75}$$

Since $f(0)=0$ and the integrand in Eq. (75) has a singularity at $\phi=0$, the finite or infinite values of the integral and, consequently, $\xi_{0t}$ depend on the behavior of the function $f(\phi)$ at $\phi \to 0$. Since $f(0)=0$ and, for infinitesimal $\phi$, we generally have $f(\phi)=\phi^\alpha$,

$$\int_1^{\phi \to 0} \frac{d\phi^*}{\sqrt{f(\phi^*)}} = \begin{cases} \lim_{\phi \to 0} \frac{1-\phi^{1-0.5\alpha}}{1-0.5\alpha} = \frac{1}{1-0.5\alpha} & \alpha < 2; \\ \lim_{\phi \to 0} \frac{1-\phi^{1-0.5\alpha}}{1-0.5\alpha} = \infty & \alpha > 2; \\ \int_1^{\phi \to 0} \frac{d\phi^*}{\sqrt{f(\phi^*)}} = -\lim_{\phi \to 0}\ln\phi = \infty & \alpha = 2. \end{cases} \tag{76}$$

Thus, if we have $\phi^\alpha$ for infinitesimal $\phi$,

$$\begin{cases} \xi_{0t} = \infty & \alpha \geq 2; \\ \xi_{0t} \text{ is finite} & \alpha < 2. \end{cases} \tag{77}$$

Note that, for $k=\infty$, we have $\alpha=2$, and for finite $k$, we have $\alpha=1$. Finite $\xi_{0t}$ is physical and practical because it allows one to avoid the damage in the entire region in the numerical solution for localized cracks. This condition was met for gradient damage models in Pham and Marigo, (2013) and is one of the reasons which for the current model $\alpha=1$ was chosen. This condition was not listed in Levitas et al. (2018) and is formulated for the first time here.

*Calibration of β.* Due to the definition of surface energy (see Eq. (19)), $\beta$ can be related to $\gamma$ and $d$:

$$\gamma = 0.5\int_{-\infty}^{+\infty}\rho_0(\psi^c+\psi^\nabla)d\xi_0 = \int_{-\xi_{0t}}^{+\xi_{0t}}\rho_0\psi^c d\xi_0 = A\int_{-\xi_{0t}}^{+\xi_{0t}}fd\xi_0 = 2A\int_0^1 f\frac{d\xi_0}{d\phi}d\phi = \sqrt{2A\beta}\int_0^1\sqrt{f}d\phi =$$

$$\sqrt{2A\beta/(k+1)}\int_0^1\sqrt{k\phi^2+\phi}d\phi \Rightarrow \beta = \frac{16(k^2+k)}{\left(2(2k+1)\sqrt{k^2+k}-\ln(2\sqrt{k^2+k}+2k+1)\right)^2}\gamma d. \tag{78}$$

Eq. (70) was utilized in derivations and Eq. (68) was used for $A$. Note that the crack has two surfaces, explaining the factor of 0.5 in the definition Eq. (19). This is different from what is presented in some other phase field models for fracture (Borden et al., 2016; Miehe and Schänzel, 2014; Weinberg and Hesch, 2017). Using Eqs. (73) and (78) we obtain for the especial cases of $k=0$: $\beta=9\gamma d/16=0.5625\gamma d$; $\xi_{0t}=0.75d$, $k=1$: $\beta=0.708\gamma d$; $\xi_{0t}=1.05d$, and $k=\infty$: $\beta=\gamma d$ and ; $\xi_{0t}=\infty$.

We would like to make an important point as to why the free energy terms $\rho_0 \psi^c$ and $\rho_0 \psi^\nabla$ should be expressed per unit undeformed volume. The damage distribution $\phi$ and the cohesion interpolation function $f(\phi)$, in the undeformed and deformed configurations, are represented in Fig. 8a and Fig. 8b, respectively.

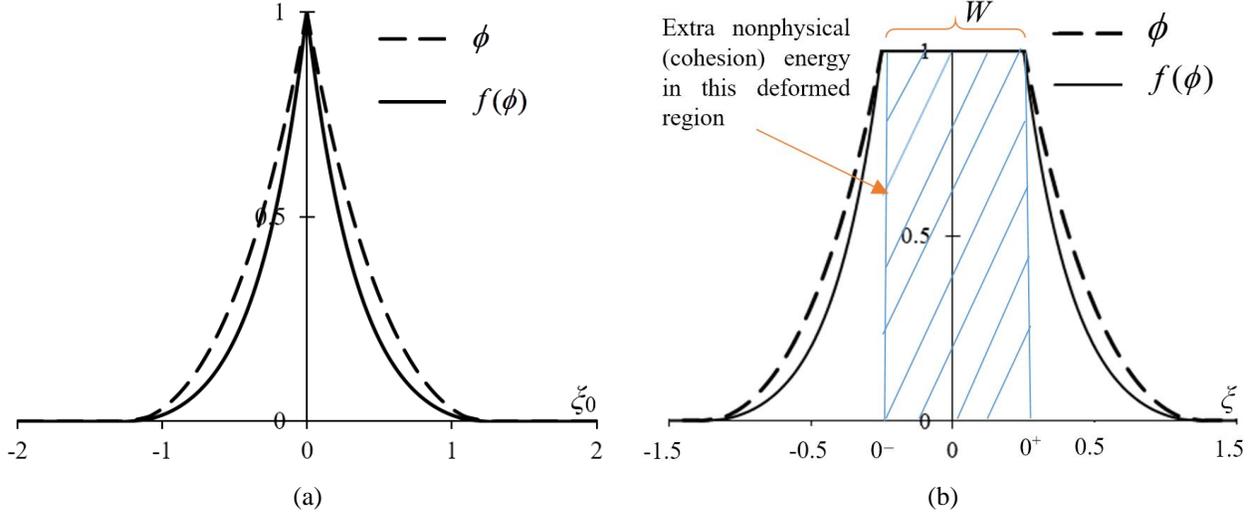

**Fig. 8.** Schematic crack profile (dashed lines) and the plot of the interpolation function $f$ (solid lines) a) for the current model, with the energy defined per unit undeformed volume, and b) if the energy is defined per unit deformed volume.

In the empty region of width $W$ between the two crack surfaces, $\phi = 1$ and $f(\phi) = 1$. Therefore, the expression of the cohesion energy, if defined per unit deformed volume, adds an extra nonphysical cohesion energy per unit current crack area $\rho_0 \psi^c(1)W = AW$ in the region between the two separated planes (see Fig. 8b). This violates an energy balance, which is why we did not use the cohesion energy per unit deformed volume and could not use the approach for introducing the surface stresses developed in (Levitas, 2013b; Levitas, 2014; Levitas and Javanbakht, 2010; Levitas and Warren, 2016) based on the deformed configuration. Our current approach based on including the term $dS/dS_0$ in Eq. (18) also involves the deformed state in term of $dS$. However, this deformation is along the crack surfaces and does not involve deformation producing the empty space.

## 10. Analytical expression for surface stresses

Inserting Eq. (73) and Eq. (78) into Eq. (72) leads to a more convenient form of the surface profile for the current model and the particular case of $k=1$, where $\beta=0.708\gamma d$ and $\xi_{0t}=1.05d$:

$$\sqrt{\phi} + \sqrt{\phi+1} = (1+\sqrt{2})e^{-\frac{|\xi_0|}{1.2d}} \quad |\xi_0| < 1.05d;$$
$$\phi = 0 \quad |\xi_0| \geq 1.05d. \quad (79)$$

Based on Eq. (29), and using Eqs. (51), (68), and (70), we obtain

$$\sigma^{st} = \rho \frac{dS}{dS_0} \frac{4\gamma}{d} \frac{k\phi^2 + \phi}{k+1}. \tag{80}$$

A plot of $\bar{\sigma}^{st} := \frac{\sigma^{st}}{4\rho\gamma dS/dS_0} = \frac{k\phi^2 + \phi}{(k+1)d}$ for $k=1$ with $\phi(\xi_0)$ from Eq. (79) for several surface width parameter $d$ is presented in Fig. 9. The maximum magnitude of $\bar{\sigma}^{st}$ is $1/d$ at $x_0=0$ (which is the same for any $k$).

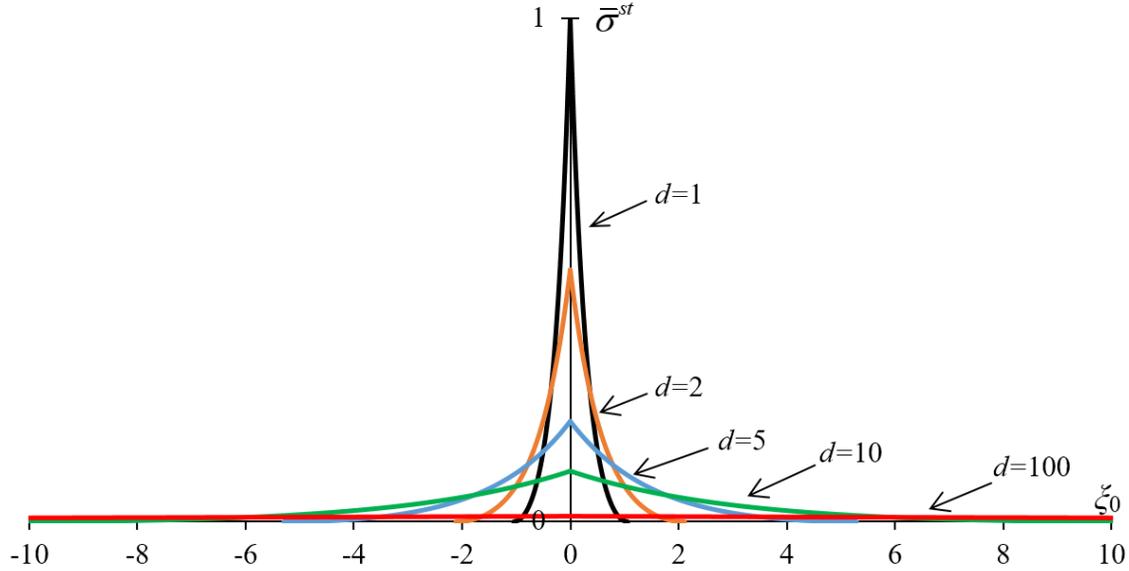

**Fig. 9.** Distribution of the surface stress $\bar{\sigma}^{st}$ for several interface widths $d$ shown near curves and $k=1$.

Since all the curves in Fig. 9 correspond to the same surface energy, the resultant surface force, which is equal to the surface energy and proportional to the area below the curves, is the same. It is clear that in all the curves $\bar{\sigma}^{st}d$ coincide. As it is shown in Eq. (73), for the same surface energy, the surface width $\xi_{0t}$ is proportional to $d$.

It was demonstrated in Levitas et al. (2018) that the effect of the surface stresses on the stress field of the crack tip is important for nanoscale $d$ and negligible for a sufficiently large $d$. Even without the surface stresses the local elastic stress field near the crack tip strongly depends on $d$ (Levitas et al. 2018) but the surface stresses decay faster ($\propto 1/d$) than the elastic stresses ($\propto 1/\sqrt{d}$), as $d$ increases.

### 11. Complete system of the equations

The final system of equations is collected below.

#### 11.1. Kinematics

Large strains

$$F = I + \nabla_{\circ} u \, ; \quad E = 0.5\left(F^T \cdot F - I\right). \tag{81}$$

Small strains

$$\varepsilon = 0.5\left(\nabla u + \nabla u^T\right). \tag{82}$$

## 11.2. Helmholtz free energy per unit mass and its contributions

$$\psi = \frac{dS}{dS_0}(\psi^{\nabla} + \psi^c) + \psi^e; \tag{83}$$

$$\rho_0 \psi^{\nabla} = \frac{1}{2}\beta(\nabla_{\circ}\phi)^2 \, ; \quad \rho_0 \psi^c = A f(\phi); \quad f(\phi) = \frac{k\phi^2 + \phi}{k+1}. \tag{84}$$

Large strains

$$\frac{dS}{dS_0} = J\left|F^{-T} \cdot m_0\right|; \quad m_0 = \nabla_{\circ}\phi / \left|\nabla_{\circ}\phi\right|; \tag{85}$$

$$\rho_0 \psi^e = (1-\phi)^n \left(\frac{1}{2} E : C_2 : E + \frac{1}{3!}(E : C_3 : E) : E + ...\right). \tag{86}$$

Small strains

$$\frac{dS}{dS_0} = 1 + \left(I - m_0 \otimes m_0\right) : \varepsilon; \quad m_0 = \nabla_{\circ}\phi / \left|\nabla_{\circ}\phi\right|; \tag{87}$$

$$\rho_0 \psi^e = (1-\phi)^n \left(\frac{1}{2}\varepsilon : C_2 : \varepsilon + \frac{1}{3!}(\varepsilon : C_3 : \varepsilon) : \varepsilon + ...\right); \tag{88}$$

## 11.3. Phase field parameters

Damage distribution

$$\begin{cases} \sqrt{k\phi} + \sqrt{k\phi + 1} = (\sqrt{k} + \sqrt{k+1})e^{-\sqrt{\frac{k\gamma}{(k+1)\beta d}}|\xi_0|} & |\xi_0| < \xi_{0t}; \\ \phi = 0 & |\xi_0| \geq \xi_{0t}, \end{cases} \tag{89}$$

Gradient and cohesive energy coefficients

$$\beta = \frac{16(k^2 + k)}{\left(2(2k+1)\sqrt{k^2 + k} - \ln(2\sqrt{k^2 + k} + 2k + 1)\right)^2} \gamma d \, ; \qquad A = \frac{2\gamma}{d}. \tag{90}$$

Location of the furthest damaged plane

$$\xi_{0t} = \frac{4(k+1)\ln(\sqrt{k} + \sqrt{k+1})}{2(2k+1)\sqrt{k^2 + k} - \ln(2\sqrt{k^2 + k} + 2k + 1)} d. \tag{91}$$

Elastic energy at damage initiation

$$\Psi_i^e = A \frac{dS^*}{dS_0^*} \frac{1}{(k+1)n}. \tag{92}$$

### 11.4. Stress tensor
Large strains

$$\begin{aligned}
&\boldsymbol{P} = \boldsymbol{P}^e + \boldsymbol{P}^{st}; \\
&\boldsymbol{P}^e = (1-\phi)^n \boldsymbol{F} \cdot \left( \boldsymbol{C}_2 : \boldsymbol{E} + \frac{1}{2} \boldsymbol{E} : \boldsymbol{C}_3 : \boldsymbol{E} + ... \right); \\
&\boldsymbol{P}^{st} = \rho_0 \frac{dS}{dS_0} (\psi^\nabla + \psi^c)(\boldsymbol{I} - \boldsymbol{m} \otimes \boldsymbol{m}) \cdot \boldsymbol{F}^{-T}; \quad \boldsymbol{m} = \nabla \phi / |\nabla \phi|; \\
&\boldsymbol{\sigma} = \boldsymbol{\sigma}_e + \boldsymbol{\sigma}_{st}; \\
&\boldsymbol{\sigma}_e = \frac{1}{J}(1-\phi)^n \boldsymbol{F} \cdot \left( \boldsymbol{C}_2 : \boldsymbol{E} + \frac{1}{2}(\boldsymbol{E} : \boldsymbol{C}_3 : \boldsymbol{E}) + ... \right) \cdot \boldsymbol{F}^T; \\
&\boldsymbol{\sigma}^{st} = \rho \frac{dS}{dS_0} (\psi^\nabla + \psi^c)(\boldsymbol{I} - \boldsymbol{m} \otimes \boldsymbol{m}).
\end{aligned} \tag{93}$$

Small strains

$$\begin{aligned}
&\boldsymbol{\sigma} = \boldsymbol{\sigma}_e + \boldsymbol{\sigma}_{st}; \\
&\boldsymbol{\sigma}_e = (1-\phi)^n \left( \boldsymbol{C}_2 : \boldsymbol{\varepsilon} + \frac{1}{2} \boldsymbol{\varepsilon} : \boldsymbol{C}_3 : \boldsymbol{\varepsilon} + ... \right); \\
&\boldsymbol{\sigma}^{st} = \rho(\psi^\nabla + \psi^c)(\boldsymbol{I} - \boldsymbol{m} \otimes \boldsymbol{m}); \quad \boldsymbol{m} = \nabla \phi / |\nabla \phi|;.
\end{aligned} \tag{94}$$

### 11.5. Ginzburg-Landau equation
*11.5.1.* Reference configuration
Compact form

$$\begin{aligned}
&\dot{\phi}(\boldsymbol{r}_0, t) = L \left[ -\frac{\partial \psi}{\partial \phi} + \nabla_\circ \cdot \frac{\partial \left( (dS/dS_0)(\psi^\nabla + \psi^c) \right)}{\partial \nabla_\circ \phi} \right]; \\
&L = \rho_0 \bar{L} = \begin{cases} 0 & \boldsymbol{m}_0 \cdot \boldsymbol{P} \cdot \boldsymbol{m}_0 \leq 0; \\ L_t & \text{otherwise.} \end{cases}
\end{aligned} \tag{95}$$

Detailed form (large strains)

$$\frac{1}{\bar{L}} \dot{\phi}(\boldsymbol{r}_0, t) + I'(\phi) \Psi^e + A \frac{dS}{dS_0} f'(\phi) = \beta \frac{dS}{dS_0} \nabla_\circ^2 \phi +$$

$$J \beta \nabla_\circ \phi \cdot \left( \boldsymbol{m}_0 \cdot \nabla_\circ \boldsymbol{F}^{-1} \cdot \frac{\boldsymbol{F}^{-T} \cdot \boldsymbol{m}_0}{|\boldsymbol{F}^{-T} \cdot \boldsymbol{m}_0|} + |\boldsymbol{F}^{-T} \cdot \boldsymbol{m}_0| \boldsymbol{F}^{-T} : \nabla_\circ \boldsymbol{F} \right) + J(\psi^\nabla + \psi^c) \times \tag{96}$$

$$\left\{ \left( \nabla_\circ \boldsymbol{m}_0 \cdot \boldsymbol{F}^{-1} \cdot \boldsymbol{F}^{-T} + \boldsymbol{m}_0 \cdot \nabla_\circ \boldsymbol{F}^{-1} \cdot \boldsymbol{F}^{-T} + \boldsymbol{m}_0 \cdot \boldsymbol{F}^{-1} \cdot \nabla_\circ \boldsymbol{F}^{-T} \right) : \frac{\boldsymbol{I} - \boldsymbol{m}_0 \otimes \boldsymbol{m}_0}{|\boldsymbol{F}^{-T} \cdot \nabla_\circ \phi|} - \right.$$

$$\frac{\nabla_{\circ} \cdot \boldsymbol{m}_0}{\left|\boldsymbol{F}^{-T} \cdot \nabla_{\circ} \phi\right|} \boldsymbol{m}_0 \cdot \boldsymbol{F}^{-1} \cdot \boldsymbol{F}^{-T} \cdot \boldsymbol{m}_0 -$$

$$\frac{\boldsymbol{F}^{-T} \cdot \nabla_{\circ} \phi}{(\boldsymbol{F}^{-T} \cdot \nabla_{\circ} \phi)^2} \cdot \left(\nabla_{\circ} \boldsymbol{F}^{-T} \cdot \nabla_{\circ} \phi + \boldsymbol{F}^{-T} \cdot \nabla_{\circ} \nabla_{\circ} \phi\right) \cdot \left(\boldsymbol{I} - \boldsymbol{m}_0 \otimes \boldsymbol{m}_0\right) \cdot \boldsymbol{F}^{-1} \cdot \boldsymbol{F}^{-T} \cdot \boldsymbol{m}_0 \bigg\} +$$

$$J \boldsymbol{m}_0 \cdot \boldsymbol{F}^{-1} \cdot \boldsymbol{F}^{-T} \cdot \frac{\boldsymbol{I} - \boldsymbol{m}_0 \otimes \boldsymbol{m}_0}{\left|\boldsymbol{F}^{-T} \cdot \nabla_{\circ} \phi\right|} \cdot \left(\left(\boldsymbol{F}^{-T} : \nabla_{\circ} \boldsymbol{F}\right)(\psi^{\nabla} + \psi^c) + \beta \nabla_{\circ} \nabla_{\circ} \phi \cdot \nabla_{\circ} \phi\right).$$

Detailed form (small strains)

$$\frac{1}{\bar{L}} \dot{\phi}(\boldsymbol{r}_0, t) + I'(\phi) \Psi^e + A\left(1 + \left(\boldsymbol{I} - \boldsymbol{m}_0 \otimes \boldsymbol{m}_0\right) : \boldsymbol{\varepsilon}\right) f'(\phi) =$$

$$\beta \left(1 + \left(\boldsymbol{I} - \boldsymbol{m}_0 \otimes \boldsymbol{m}_0\right) : \boldsymbol{\varepsilon}\right) \nabla_{\circ}^2 \phi + \beta \nabla_{\circ} \phi \cdot \nabla_{\circ} \boldsymbol{\varepsilon} : \left(\boldsymbol{I} - \boldsymbol{m}_0 \otimes \boldsymbol{m}_0\right) -$$

$$\beta \left(1 + \frac{\psi^c}{\psi^{\nabla}}\right) \left(\left|\nabla_{\circ} \phi\right| \nabla_{\circ} \boldsymbol{m}_0 : \left(\boldsymbol{\varepsilon} \cdot \left(\boldsymbol{I} - \boldsymbol{m}_0 \otimes \boldsymbol{m}_0\right)\right)\right) + \nabla_{\circ} \phi \cdot \nabla_{\circ} \boldsymbol{\varepsilon} : \left(\boldsymbol{I} - \boldsymbol{m}_0 \otimes \boldsymbol{m}_0\right) + \quad (97)$$

$$\boldsymbol{m}_0 \cdot \boldsymbol{\varepsilon} \cdot \left(\boldsymbol{m}_0 \nabla_{\circ}^2 \phi - \boldsymbol{m}_0 \cdot (\nabla_{\circ} \nabla_{\circ} \phi)^T\right) - \nabla_{\circ} \phi \cdot \boldsymbol{\varepsilon} \cdot \left(\boldsymbol{I} - \boldsymbol{m}_0 \otimes \boldsymbol{m}_0\right) \cdot \left(\boldsymbol{\varepsilon} : \nabla_{\circ} \boldsymbol{\varepsilon}\right) -$$

$$2\beta \boldsymbol{m}_0 \cdot \boldsymbol{\varepsilon} \cdot \left(\boldsymbol{I} - \boldsymbol{m}_0 \otimes \boldsymbol{m}_0\right) \cdot \nabla_{\circ} \nabla_{\circ} \phi \cdot \boldsymbol{m}_0,$$

*11.5.2.* Deformed configuration

$$\frac{D\phi(\boldsymbol{r},t)}{Dt} = \frac{\partial \phi(\boldsymbol{r},t)}{\partial t} + \boldsymbol{v} \cdot \nabla \phi = L\left[-\frac{\partial \psi}{\partial \phi} + \nabla\left(\boldsymbol{F}^{-1} \cdot \frac{\partial \left((dS/dS_0)(\psi^{\nabla} + \psi^c)\right)}{\partial \nabla \phi}\right) : \boldsymbol{F}\right];$$

$$L = \rho_0 \bar{L} = \begin{cases} 0 & \boldsymbol{m} \cdot \boldsymbol{\sigma} \cdot \boldsymbol{m} \leq 0; \\ L_t & \text{otherwise.} \end{cases} \quad (98)$$

### 11.6. Momentum balance equation

Large strains

$$\nabla_{\circ} \cdot \boldsymbol{P} + \rho_0 \boldsymbol{f} = \rho_0 \dot{\boldsymbol{v}}. \quad (99)$$

Small strains

$$\nabla \cdot \boldsymbol{\sigma} + \rho \boldsymbol{f} = \rho \dot{\boldsymbol{v}}. \quad (100)$$

### 11.7. Boundary condition for $\phi$

Reference configuration

$$\boldsymbol{n}_0 \cdot \frac{\partial \psi}{\partial \nabla_{\circ} \phi} = 0, \quad \text{or} \quad \phi = 1 \text{ or } \phi = 0. \quad (101)$$

Deformed configuration

$$\boldsymbol{n} \cdot \frac{\partial \psi}{\partial \nabla \phi} = 0, \quad \text{or} \quad \phi = 1 \text{ or } \phi = 0. \quad (102)$$

## 12. Parameter calibration

In this section, we calibrate the phase field parameters and the stress-strain curves for homogenous uniaxial tension. The Cauchy stress $\sigma_1$ vs. the engineering strain $e_1$ curves along the crystallographic (and loading) directions <100> and <111> of the perfect crystal of silicon (Si) with a diamond lattice are taken from the first principle simulations in Černý, Řehák, Umeno, & Pokluda (2012). Our equations must be transformed into the stress and strain measures presented in (Černý et al., 2012). For stresses, Eq. (66) is used and simplified to

$$\sigma_1 = P_1/(1+e_2+e_3) \quad \text{and} \quad \sigma_1 = (1+e_1)T_1/(1+e_2+e_3). \tag{103}$$

Components of the engineering and Lagrangian strains are related as

$$E_i = e_i + 0.5 e_i^2. \tag{104}$$

As Eq. (103) shows, the Cauchy stress depends on the other components of the strain tensor as well. Thus, $e_2$ and $e_3$ must be evaluated to obtain $\sigma_1$. We use the known elastic constants $C_{11}=C_{22}=C_{33}=166$, $C_{12}=C_{13}=C_{23}=63$, $C_{44}=C_{55}=C_{66}=80$ GPa (Hennig, Wadehra, Driver, Parker, Umrigar, & Wilkins, 2010) with axis 1 along the direction [100]. By rotating the elastic matrix, we obtain the elastic constants for the coordinate system with axis 1 along direction [111]: $C_{11}=204.043$, $C_{22}=194.543$, $C_{33}=194.543$, $C_{12}=43.968$, $C_{23}=53.435$, $C_{13}=44.054$, $C_{44}=60.957$, $C_{55}=70.5$, $C_{66}=61.0$ GPa. Inverting the elastic matrices leads to compliance, and for the case of only one nonzero component $T_1 \neq 0$, we obtain

$$E_1 = 0.0076 T_1/(1-\phi)^n, \; E_2 = E_3 = -0.0021 T_1/(1-\phi)^n \text{ for [100] and}$$
$$E_1 = 0.0053 T_1/(1-\phi)^n, \; E_2 = E_3 = -0.0009 T_1/(1-\phi)^n \text{ for [111]}, \tag{105}$$

where the stresses are in GPa. To plot the $\sigma$-$e$ curves, we combine Eqs. (104), (105), and (49), exclude the order parameter and resolve for $\sigma$ (for convenience, the superscript 1 is eliminated from here on). The obtained $\sigma$-$e$ curve is shown in Fig. 10.

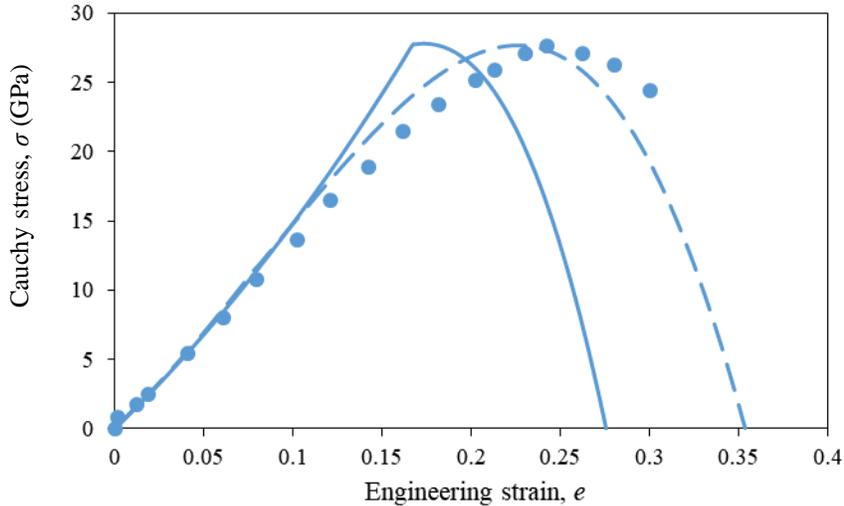

(a)

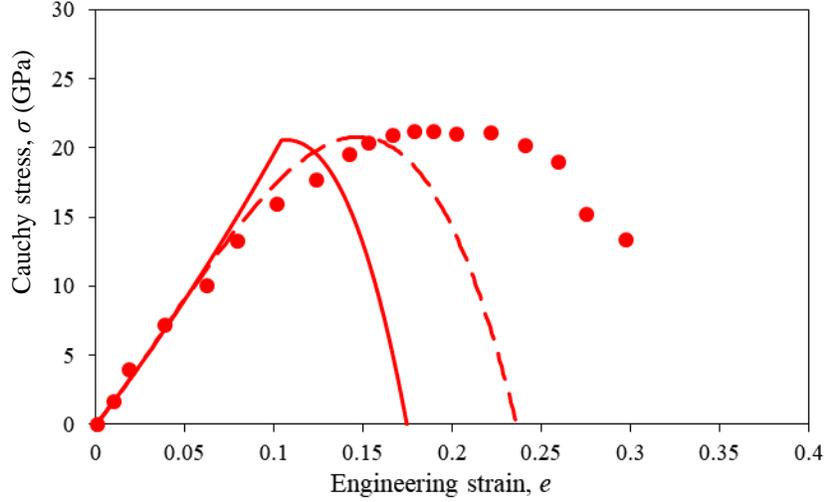

(b)

**Fig. 10.** Cauchy stress-engineering strain diagrams for the homogeneous phase field solution for loading in the a) [111] and b) [100] directions. Eq. (105) are utilized. Solid lines and dashed lines correspond to $k=1$ and $k=15$, respectively. Points represent the first principle simulation results from (Černý et al., 2012)).

We vary parameter $A$ to achieve the best fit and use $n=1$ and two different values of $k=1$ and 15. It is clear that $k=15$ gives a better correspondence with the first principle simulations in (Černý et al., 2012), but there is still room for improvement. The main reason for the discrepancy is related to our desire to keep the theory simple and use the linear relationship between $T$ and $E$ before damage starts. However, if we sacrifice the accuracy for small strains and change $C_2$ to 110 GPa for [100] and 121 GPa for [111], i.e., change Eqs. (105) to

$$E_1=0.0091\, T_1/(1-\phi)^n,\ E_2=E_3=-0.0021 T_1/(1-\phi)^n \text{ for [100] and}$$
$$E_1=0.0083 T_1/(1-\phi)^n,\ E_2=E_3=-0.0009 T_1/(1-\phi)^n \text{ for [111],}$$

(106)

we obtain a much better description of the larger strains (Fig. 11), especially for $k=15$.

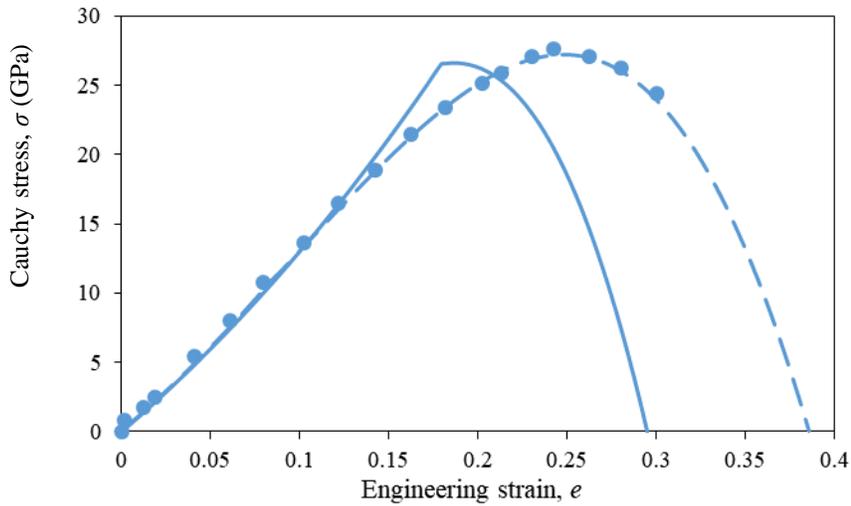

(a)

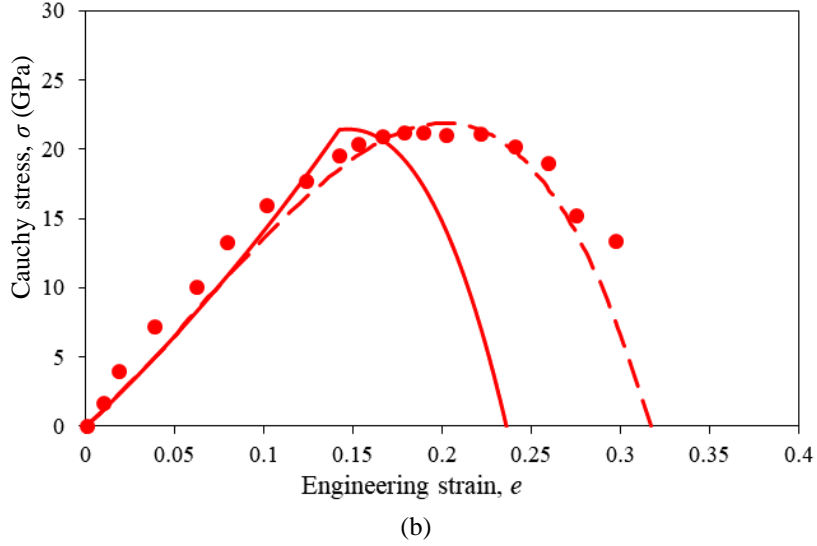

(b)

**Fig. 11.** Cauchy stress-engineering strain diagrams for the homogeneous phase field solution for loading in the a) [111] and b) [100] directions. Eq. (106) are utilized. Solid lines and dashed lines correspond to $k=1$ and $k=15$, respectively. Points represent the first principle simulation results from (Černý et al., 2012).

All the parameters for our phase field are presented in Table 1 which are based on the dashed lines, i.e., $k=15$ in Fig. 11.

**Table 1**
Phase field parameters for Si for tension in the [100] and [111] directions. The references are shown in front of each parameter.

| Crystallographic directions | [100] | [111] |
|---|---|---|
| $n$ (best fit) | 1 | 1 |
| $C_2$ (GPa) (best fit) | 110 | 121 |
| $A$ (GPa) (best fit) | 6.0 | 4.2 |
| $\gamma$ (J/m$^2$) (Messmer and Bilello, 1981) | 1.34 | 1.14 |
| $d$ (nm) [current model, Eq. (68)] | 0.45 | 0.54 |
| $d_0$ (nm) [from Eq. (107)] | 0.54 | 0.31 |
| $N$ (Levitas et al., 2018) | 0.83 | 1.7 |
| $\beta \times 10^9$ (N) [current model, Eq. (78)] | 0.57 | 0.58 |
| $\xi_{0t}$ (nm) [current model, Eq. (73)] | 0.93 | 1.1 |

Knowing $A$, $C_2$, and the surface energy $\gamma$, we can obtain $d$ from Eq. (68). The gradient coefficient is then determined from Eq. (78) as $\beta=0.938\gamma d$ and $\xi_{0t}=2.06d$ is obtained from Eq. (73). The distance between two <100> and <111> planes $d_0$ is given as

$$a=d_{0,100}=0.539 \text{ nm (Černý et al., 2012)}; \quad d_{0,111}= a/\sqrt{3}=0.311 \text{ nm}, \tag{107}$$

where $a$ is the lattice constant. The value $N:=d/d_0$ reported in Table 1 is close to 1, as would be expected for a nanoscale model, i.e., the distance between two crack surfaces is equal to the distance between the two nearest atomic planes (Levitas et al., 2018). The small discrepancy appears because some data (e.g., surface energy) are taken from different studies and due to use of the second-order elasticity. Therefore, surface stresses cannot be neglected for this material with small values of *N*.

## 13. Concluding remarks

A thermodynamically consistent phase field approach to the fracture in large strain framework is presented. One of the main contributions is the introduction of the surface stresses, which requires large-strain formulation even for infinitesimal strains. Another necessity of the large-strain formulation is to avoid artificial penalization of the cohesive energy in the space between crack surfaces. Therefore, cohesive and gradient energies should be defined in the reference configuration, and only strains along the crack surface are allowable in their expressions. In solid surfaces, the surface stresses consist of the elastic and structural parts. The elastic contribution to the surface stresses results automatically from the solution of the coupled Ginzburg-Landau and mechanics equations. After comparing the total surface stresses from the model with experiments or atomistic simulations, one can develop a more sophisticated constitutive development for the elastic surface stresses, if necessary. Thus, the main focus is on the structural part, which also exists within the liquid-liquid and liquid-gas interfaces. The critical point is that a physical phenomenon such as surface stresses is resulted by utilizing geometric nonlinearities. Thus, the gradient and the cohesion energies are multiplied by the ratio of areas of elemental crack surfaces after deformation and before it, $dS/dS_0$. This leads to the desired isotropic biaxial surface tension, with the force per unit length equal to the surface energy per unit deformed surface. The explicit expression for the damage evolution equation has been obtained for the fully geometrically nonlinear formulation, which leads to significant complication of the equation. Several approximate expressions for the Ginzburg-Landau equation under different geometric simplifications are presented. Without introducing some geometrically nonlinear terms, we could not introduce consistent surface stresses, even at small strains. This highlights the necessity of starting with a fully geometrically nonlinear formulation even for small strains. The surface stresses affect the driving force of fracture in two ways. First, they disturb the mechanical equilibrium equation by means of the additional contribution to the stress field which arises from the surface stresses. In an other words, the mechanical equilibrium equation is changed from $\nabla \cdot \boldsymbol{\sigma}^e + \rho \boldsymbol{f} = \rho \dot{\boldsymbol{v}}$ ($\nabla_\circ \cdot \boldsymbol{P}^e + \rho_0 \boldsymbol{f} = \rho_0 \dot{\boldsymbol{v}}$) to $\nabla \cdot (\boldsymbol{\sigma}^e + \boldsymbol{\sigma}^{st}) + \rho \boldsymbol{f} = \rho \dot{\boldsymbol{v}}$ ($\nabla_\circ \cdot (\boldsymbol{P}^e + \boldsymbol{P}^{st}) + \rho_0 \boldsymbol{f} = \rho_0 \dot{\boldsymbol{v}}$). Second, it leads to several additional sophisticated terms in the Ginzburg-Landau equation. At the level of the specific models:

1) A more general degradation function was introduced, including an additional material parameter *n* in comparison to the traditional physical or fitting parameters in any phase field approach to fracture. This parameter allows an improved description of the local stress-stress curve at the nanoscale when it is known from atomistic simulations.

2) The interpolation function for the cohesive energy introduced in this model leads to the crack surface with a finite width of the damaged zone. This is in contrast to all the previous phase field models, in which the intact phase is mathematically located at infinity.

3) The damage initiation criterion is formulated in the current model. Parameters *n* and *k* produce portioning of the total stress work into the elastic work before damage initiation and during damage.

Implementation of the derived equations for the surface stresses and the Ginzburg-Landau equation into a finite element code and the solution of some boundary-value problems will be considered in the next paper. Generalization for the anisotropic surface energy (Clayton and Knap, 2015; Hakim and Karma, 2009; Mozaffari and Voyiadjis, 2015) can also be performed in future work. The most popular method is to consider the anisotropic gradient energy (Wheeler and McFadden, 1997) in addition to the anisotropic cohesion energy, in which the peak stress for homogenous states is anisotropic as well. The other generalization of the developed theory can be performed for the interaction between crack propagation and phase transformation (Mamivand, Asle Zaeem, & El Kadiri, 2014), plasticity (Mozaffari and Voyiadjis, 2016; Ruffini and Finel, 2015).

Note that Levitas et al. (2018) developed a phase field model for fracture, which is valid for an arbitrary scale, from nano to macro. Development of the current paper can be implemented for that model as well. However, the contribution of the surface stresses is essential at the nanoscale only.

**Appendixes. Some derivations**

- Here we obtain explicit expression for $\dfrac{d}{d\boldsymbol{F}}\dfrac{dS}{dS_0}$ in Eq. (27) is such way

$$\dfrac{d}{d\boldsymbol{F}}\dfrac{dS}{dS_0} = \dfrac{d}{d\boldsymbol{F}}(J|\boldsymbol{F}^{-T}\cdot\boldsymbol{m}_0|) = \dfrac{dJ}{d\boldsymbol{F}}|\boldsymbol{F}^{-T}\cdot\boldsymbol{m}_0| + J\dfrac{d}{d\boldsymbol{F}}|\boldsymbol{F}^{-T}\cdot\boldsymbol{m}_0| = J\boldsymbol{F}^{-T}|\boldsymbol{F}^{-T}\cdot\boldsymbol{m}_0| - \\ J|\boldsymbol{F}^{-T}\cdot\boldsymbol{m}_0|(\boldsymbol{m}\otimes\boldsymbol{m})\cdot\boldsymbol{F}^{-T} = J|\boldsymbol{F}^{-T}\cdot\boldsymbol{m}_0|(\boldsymbol{I}-\boldsymbol{m}\otimes\boldsymbol{m})\cdot\boldsymbol{F}^{-T} = \dfrac{dS}{dS_0}(\boldsymbol{I}-\boldsymbol{m}\otimes\boldsymbol{m})\cdot\boldsymbol{F}^{-T},$$ (108)

where the equality $dJ/d\boldsymbol{F} = J\boldsymbol{F}^{-T}$ was used, and the second derivative was manipulated by defining

$$\boldsymbol{Z} := \dfrac{d}{d\boldsymbol{F}}|\boldsymbol{F}^{-T}\cdot\boldsymbol{m}_0| = \dfrac{d}{d\boldsymbol{F}}\left((\boldsymbol{F}^{-T}\cdot\boldsymbol{m}_0)\cdot(\boldsymbol{m}_0\cdot\boldsymbol{F}^{-1})\right)^{1/2}.$$ (109)

In the component form

$$Z_{mn} = \frac{d}{dF_{mn}}\left(F_{ij}^{-T}m_{0j}m_{0k}F_{ki}^{-1}\right)^{1/2} = \frac{1}{2\left|\boldsymbol{F}^{-T}\cdot\boldsymbol{m}_0\right|}\frac{d}{dF_{mn}}\left(F_{ji}^{-1}m_{0j}m_{0k}F_{ik}^{-T}\right) =$$
$$\frac{1}{\left|\boldsymbol{F}^{-T}\cdot\boldsymbol{m}_0\right|}\frac{dF_{ji}^{-1}}{dF_{mn}}m_{0j}m_{0k}F_{ik}^{-T}.$$
(110)

Now we use $dF_{ji}^{-1}/dF_{mn} = -F_{jm}^{-1}F_{ni}^{-1}$ which is a consequence of the equations $\boldsymbol{F}^{-1}\cdot\boldsymbol{F} = \boldsymbol{I}$, $d\boldsymbol{F}^{-1}\cdot\boldsymbol{F} + \boldsymbol{F}^{-1}\cdot d\boldsymbol{F} = \boldsymbol{0}$, and $d\boldsymbol{F}^{-1} = -\boldsymbol{F}^{-1}\cdot d\boldsymbol{F}\cdot\boldsymbol{F}$. Then we continue from the last term in Eq. (109):

$$Z_{mn} = -\frac{1}{\left|\boldsymbol{F}^{-T}\cdot\boldsymbol{m}_0\right|}F_{jm}^{-1}F_{ni}^{-1}m_{0j}m_{0k}F_{ik}^{-T} = -\frac{1}{\left|\boldsymbol{F}^{-T}\cdot\boldsymbol{m}_0\right|}F_{mj}^{-T}m_{0j}F_{ik}^{-T}m_{0k}F_{ni}^{-1} =$$
$$-\frac{1}{\left|\boldsymbol{F}^{-T}\cdot\boldsymbol{m}_0\right|}\frac{1}{(\nabla_{\!\circ}\phi)^2}\left(F_{mj}^{-T}\nabla_{\!\circ}\phi_j\right)\left(F_{ik}^{-T}\nabla_{\!\circ}\phi_k\right)F_{ni}^{-1} = -\frac{1}{\left|\boldsymbol{F}^{-T}\cdot\boldsymbol{m}_0\right|}\frac{1}{(\nabla_{\!\circ}\phi)^2}\nabla\phi_m\nabla\phi_i F_{ni}^{-1} =$$
(111)
$$-\frac{1}{\left|\boldsymbol{F}^{-T}\cdot\boldsymbol{m}_0\right|}\frac{(\nabla\phi)^2}{(\nabla_{\!\circ}\phi)^2}m_m m_i F_{ni}^{-1} = -\left|\boldsymbol{F}^{-T}\cdot\boldsymbol{m}_0\right|m_m m_i F_{in}^{-T},$$

which is the component form of

$$\boldsymbol{Z} = -\left|\boldsymbol{F}^{-T}\cdot\boldsymbol{m}_0\right|(\boldsymbol{m}\otimes\boldsymbol{m})\cdot\boldsymbol{F}^{-T}.$$
(112)

Note that we also used

$$\nabla\phi = \boldsymbol{F}^{-T}\cdot\nabla_{\!\circ}\phi, \qquad \nabla\phi_i = F_{ij}^{-T}\nabla_{\!\circ}\phi_j.$$
(113)

- Let us calculate $\boldsymbol{M}$, which is defined from Eq. (37) as

$$\boldsymbol{M} = \frac{\partial}{\partial\nabla_{\!\circ}\phi}\left|\boldsymbol{F}^{-T}\cdot\boldsymbol{m}_0\right| = \frac{\partial}{\partial\nabla_{\!\circ}\phi}\left((\boldsymbol{F}^{-T}\cdot\boldsymbol{m}_0)\cdot(\boldsymbol{m}_0\cdot\boldsymbol{F}^{-1})\right)^{1/2}.$$
(114)

In a component form

$$M_l = \frac{\partial}{\partial\nabla_{\!\circ}\phi_l}\left(F_{ij}^{-T}m_{0j}m_{0k}F_{ki}^{-1}\right)^{1/2} = \frac{1}{2\left|\boldsymbol{F}^{-T}\cdot\boldsymbol{m}_0\right|}\frac{\partial}{\partial\nabla_{\!\circ}\phi_l}\left(F_{ij}^{-T}m_{0j}m_{0k}F_{ik}^{-T}\right) =$$
$$\frac{1}{\left|\boldsymbol{F}^{-T}\cdot\boldsymbol{m}_0\right|}\left(F_{ij}^{-T}m_{0k}F_{ik}^{-T}\right)\frac{\partial m_{0j}}{\partial\nabla_{\!\circ}\phi_l} = \frac{1}{\left|\nabla_{\!\circ}\phi\right|\left|\boldsymbol{F}^{-T}\cdot\boldsymbol{m}_0\right|}m_{0k}F_{ki}^{-1}F_{ij}^{-T}\left(I_{jl}-m_{0j}m_{0l}\right),$$
(115)

where we used

$$\frac{\partial\boldsymbol{m}_0}{\partial(\nabla_{\!\circ}\phi)} = \frac{\partial\left(\nabla_{\!\circ}\phi/\left|\nabla_{\!\circ}\phi\right|\right)}{\partial\nabla_{\!\circ}\phi} = \frac{(\partial\nabla_{\!\circ}\phi/\partial\nabla_{\!\circ}\phi)\left|\nabla_{\!\circ}\phi\right| - \left(\partial\left|\nabla_{\!\circ}\phi\right|/\partial\nabla_{\!\circ}\phi\right)\otimes\nabla_{\!\circ}\phi}{(\nabla_{\!\circ}\phi)^2} =$$
(116)

$$\frac{I|\nabla_\circ\phi| - (\nabla_\circ\phi/|\nabla_\circ\phi|) \otimes \nabla_\circ\phi}{(\nabla_\circ\phi)^2} = \frac{I - m_0 \otimes m_0}{|\nabla_\circ\phi|}.$$

Finally, in index-free notations, we have

$$M = m_0 \cdot F^{-1} \cdot F^{-T} \cdot \frac{I - m_0 \otimes m_0}{|F^{-T} \cdot \nabla_\circ\phi|}. \tag{117}$$

Since $(I - m_0 \otimes m_0) \cdot m_0 = m_0 - m_0 = 0$ and, consequently $(I - m_0 \otimes m_0) \cdot \nabla_\circ\phi = 0$ thus, $M \cdot \nabla_\circ\phi = 0$ and the last term in Eq. (36) vanishes.

- $Y$ is defined in Eq. (37) as

$$Y = \nabla_\circ \frac{dS}{dS_0} = \nabla_\circ \left(J|F^{-T} \cdot m_0|\right) = J\nabla_\circ \left((F^{-T} \cdot m_0) \cdot (m_0 \cdot F^{-1})\right)^{1/2} + |F^{-T} \cdot m_0|\nabla_\circ J. \tag{118}$$

For the first term on the right side, we use

$$Y'_l := \left(\left(F_{ij}^{-T} m_{0j} m_{0k} F_{ki}^{-1}\right)^{1/2}\right)_{,l} = \frac{1}{2|F^{-T} \cdot m_0|}\left(m_{0j} F_{ji}^{-1} F_{ik}^{-T} m_{0k}\right)_{,l} =$$

$$\frac{1}{|F^{-T} \cdot m_0|}\left(m_{0j} F_{ji,l}^{-1} F_{ik}^{-T} m_{0k} + m_{0j,l} F_{ji}^{-1} F_{ik}^{-T} m_{0k}\right). \tag{119}$$

Thus, in the direct tensor notations, $Y$ is

$$Y = \frac{J}{|F^{-T} \cdot m_0|}\left((m_0 \cdot \nabla_\circ F^{-1}) \cdot (F^{-T} \cdot m_0) + \nabla_\circ m_0 \cdot F^{-1} \cdot F^{-T} \cdot m_0\right) + |F^{-T} \cdot m_0|\nabla_\circ J =$$

$$J\left(m_0 \cdot \nabla_\circ F^{-1} + \nabla_\circ m_0 \cdot F^{-1}\right) \cdot \frac{F^{-T} \cdot m_0}{|F^{-T} \cdot m_0|} + J|F^{-T} \cdot m_0|F^{-T} : \nabla_\circ F = \tag{120}$$

$$J\left(m_0 \cdot \nabla_\circ F^{-1} + \nabla_\circ m_0 \cdot F^{-1}\right) \cdot m + J|F^{-T} \cdot m_0|F^{-T} : \nabla_\circ F.$$

We have used $\nabla_\circ J = \frac{\partial J}{\partial F} : \nabla_\circ F = JF^{-T} : \nabla_\circ F$ and $\nabla_\circ m_0$ can be evaluated as

$$\nabla_\circ m_0 = \nabla_\circ \frac{\nabla_\circ\phi}{|\nabla_\circ\phi|} = \frac{\nabla_\circ\nabla_\circ\phi|\nabla_\circ\phi| - (\nabla_\circ\nabla_\circ\phi \cdot \nabla_\circ\phi/|\nabla_\circ\phi|) \otimes \nabla_\circ\phi}{(\nabla_\circ\phi)^2} = \nabla_\circ\nabla_\circ\phi \cdot \frac{I - m_0 \otimes m_0}{|\nabla_\circ\phi|}, \tag{121}$$

which is a symmetric tensor. Note that $m_0 \cdot \nabla_\circ m_0 = 0$ and therefore $\nabla_\circ\phi \cdot \nabla_\circ m_0 = 0$ while evaluating $\nabla_\circ\phi \cdot Y$ in Eq. (36).

- Scalar $N$ is defined in Eq. (37) as

$$N = \nabla_\circ \cdot \boldsymbol{M} = \left( \frac{m_{0k} F_{ki}^{-1} F_{ij}^{-T} \left( I_{jl} - m_{0j} m_{0l} \right)}{\left( F_{hm}^{-T} \nabla_\circ \phi_m F_{hn}^{-T} \nabla_\circ \phi_n \right)^{1/2}} \right)_{,l} =$$

$$\left( m_{0k,l} F_{ki}^{-1} F_{ij}^{-T} + m_{0k} F_{ki,l}^{-1} F_{ij}^{-T} + m_{0k} F_{ki}^{-1} F_{ij,l}^{-T} \right) \frac{I_{jl} - m_{0j} m_{0l}}{\left| \boldsymbol{F}^{-T} \cdot \nabla_\circ \phi \right|} - m_{0k} F_{ki}^{-1} F_{ij}^{-T} \frac{m_{0j,l} m_{0l} + m_{0j} m_{0l,l}}{\left| \boldsymbol{F}^{-T} \cdot \nabla_\circ \phi \right|} - \quad (122)$$

$$\frac{F_{hm}^{-T} \nabla_\circ \phi_m}{\left( \boldsymbol{F}^{-T} \cdot \nabla_\circ \phi \right)^2} \left( F_{hn,l}^{-T} \nabla_\circ \phi_n + F_{hn}^{-T} \nabla_\circ \phi_{n,l} \right) m_{0k} F_{ki}^{-1} F_{ij}^{-T} \left( I_{jl} - m_{0j} m_{0l} \right).$$

We used $m_{0j,l} m_{0l} = 0$, i.e., $\nabla_\circ \boldsymbol{m}_0 \cdot \boldsymbol{m}_0 = \boldsymbol{0}$, which comes from $m_{0l,j} m_{0j} = m_{0j,l} m_{0j} = 0.5(m_{0j} m_{0j})_{,l} = 0$, where the symmetry property of $\nabla_\circ \boldsymbol{m}_0$ is used from Eq. (121). Thus,

$$N = \left( \nabla_\circ \boldsymbol{m}_0 \cdot \boldsymbol{F}^{-1} \cdot \boldsymbol{F}^{-T} + \boldsymbol{m}_0 \cdot \nabla_\circ \boldsymbol{F}^{-1} \cdot \boldsymbol{F}^{-T} + \boldsymbol{m}_0 \cdot \boldsymbol{F}^{-1} \cdot \nabla_\circ \boldsymbol{F}^{-T} \right) : \frac{\boldsymbol{I} - \boldsymbol{m}_0 \otimes \boldsymbol{m}_0}{\left| \boldsymbol{F}^{-T} \cdot \nabla_\circ \phi \right|} -$$

$$\frac{\nabla_\circ \cdot \boldsymbol{m}_0}{\left| \boldsymbol{F}^{-T} \cdot \nabla_\circ \phi \right|} \boldsymbol{m}_0 \cdot \boldsymbol{F}^{-1} \cdot \boldsymbol{F}^{-T} \cdot \boldsymbol{m}_0 - \quad (123)$$

$$\frac{\boldsymbol{F}^{-T} \cdot \nabla_\circ \phi}{(\boldsymbol{F}^{-T} \cdot \nabla_\circ \phi)^2} \cdot \left( \nabla_\circ \boldsymbol{F}^{-T} \cdot \nabla_\circ \phi + \boldsymbol{F}^{-T} \cdot \nabla_\circ \nabla_\circ \phi \right) \cdot \left( \boldsymbol{I} - \boldsymbol{m}_0 \otimes \boldsymbol{m}_0 \right) \cdot \boldsymbol{F}^{-1} \cdot \boldsymbol{F}^{-T} \cdot \boldsymbol{m}_0,$$

One can evaluate the scalar $\nabla_\circ \cdot \boldsymbol{m}_0 = \nabla_\circ \cdot \frac{\nabla_\circ \phi}{\left| \nabla_\circ \phi \right|} = \frac{\nabla_\circ^2 \phi - \boldsymbol{m}_0 \cdot \nabla_\circ \nabla_\circ \phi \cdot \boldsymbol{m}_0}{\left| \nabla_\circ \phi \right|}$.

- Here, we simplify Eq. (18)$_2$ and Eqs.(114), (118), and (122) for the small strains and rotations in such a manner:

$$\frac{dS}{dS_0} = J \left| \boldsymbol{F}^{-T} \cdot \boldsymbol{m}_0 \right| = J \left( (\boldsymbol{F}^{-T} \cdot \boldsymbol{m}_0) \cdot (\boldsymbol{m}_0 \cdot \boldsymbol{F}^{-1}) \right)^{1/2} = J \left[ \left( (\boldsymbol{I} - \boldsymbol{\varepsilon} - \boldsymbol{\omega}) \cdot \boldsymbol{m}_0 \right) \cdot \left( \boldsymbol{m}_0 \cdot (\boldsymbol{I} - \boldsymbol{\varepsilon} + \boldsymbol{\omega}) \right) \right]^{1/2} \simeq$$

$$J \left( 1 - 2 (\boldsymbol{m}_0 \otimes \boldsymbol{m}_0) : \boldsymbol{\varepsilon} \right)^{1/2} \simeq (1 + \boldsymbol{I} : \boldsymbol{\varepsilon}) \left( 1 - (\boldsymbol{m}_0 \otimes \boldsymbol{m}_0) : \boldsymbol{\varepsilon} \right) = 1 + (\boldsymbol{I} - \boldsymbol{m}_0 \otimes \boldsymbol{m}_0) : \boldsymbol{\varepsilon}. \quad (124)$$

We neglect the higher-order terms in $\boldsymbol{\varepsilon}$ and $\boldsymbol{\omega}$ and their combination. Using $\left| \boldsymbol{F}^{-T} \cdot \boldsymbol{m}_0 \right| \simeq 1 - (\boldsymbol{m}_0 \otimes \boldsymbol{m}_0) : \boldsymbol{\varepsilon}$ from Eq. (124), we obtain

$$\boldsymbol{M} = \frac{\partial \left| \boldsymbol{F}^{-T} \cdot \boldsymbol{m}_0 \right|}{\partial \nabla_\circ \phi} = \frac{\partial \left( 1 - (\boldsymbol{m}_0 \otimes \boldsymbol{m}_0) : \boldsymbol{\varepsilon} \right)}{\partial \nabla_\circ \phi} = -2 \boldsymbol{m}_0 \cdot \boldsymbol{\varepsilon} \cdot \frac{\boldsymbol{I} - \boldsymbol{m}_0 \otimes \boldsymbol{m}_0}{\left| \nabla_\circ \phi \right|}. \quad (125)$$

Using Eq. (124) again, we obtain

$$Y = \nabla_{\circ}(dS/dS_0) = \nabla_{\circ}\left(1+\left(I-m_0 \otimes m_0\right):\varepsilon\right) =$$
$$\nabla_{\circ}\left(\left(I-m_0 \otimes m_0\right):\varepsilon\right) = \left(I-m_0 \otimes m_0\right):\nabla_{\circ}\varepsilon - 2m_0\cdot\varepsilon\cdot\nabla_{\circ}m_0. \quad (126)$$

Then, by utilizing Eq. (125), we obtain

$$N = \nabla_{\circ}\cdot M = -2\nabla_{\circ}\cdot\left(m_0\cdot\varepsilon\cdot\frac{I-m_0 \otimes m_0}{|\nabla_{\circ}\phi|}\right) = -2\nabla_{\circ}m_0:\left(\varepsilon\cdot\frac{I-m_0 \otimes m_0}{|\nabla_{\circ}\phi|}\right) -$$
$$2m_0\cdot\nabla_{\circ}\varepsilon:\frac{I-m_0 \otimes m_0}{|\nabla_{\circ}\phi|} - 2m_0\cdot\varepsilon\cdot\left(m_0\nabla_{\circ}^2\phi - m_0\cdot(\nabla_{\circ}\nabla_{\circ}\phi)^T\right)/(\nabla_{\circ}\phi)^2. \quad (127)$$

The direct simplification of $M$, $Y$, and $N$ from Eqs. (117), (120), and (123) is given as follows. We neglect the higher-order terms in $\varepsilon$; thus, $(I-\varepsilon)\cdot(I-\varepsilon) \simeq I-2\varepsilon$. $M$ is simplified as

$$M = m_0\cdot F^{-1}\cdot F^{-T}\cdot\frac{I-m_0 \otimes m_0}{|F^{-T}\cdot\nabla_{\circ}\phi|} \simeq m_0\cdot(I-\varepsilon)\cdot(I-\varepsilon)\cdot\frac{I-m_0 \otimes m_0}{|\nabla_{\circ}\phi|} =$$
$$m_0\cdot(I-2\varepsilon)\cdot\frac{I-m_0 \otimes m_0}{|\nabla_{\circ}\phi|} = -2m_0\cdot\varepsilon\cdot\frac{I-m_0 \otimes m_0}{|\nabla_{\circ}\phi|}, \quad (128)$$

where $\left(I-m_0 \otimes m_0\right)\cdot m_0 = m_0 - m_0 = 0$ was used. Similarly, $Y$ is

$$Y = J\left(m_0\cdot\nabla_{\circ}F^{-1} + \nabla_{\circ}m_0\cdot F^{-1}\right)\cdot\frac{F^{-T}\cdot m_0}{|F^{-T}\cdot m_0|} + J\left|F^{-T}\cdot m_0\right|F^{-T}:\nabla_{\circ}F \simeq$$
$$\left(-m_0\cdot\nabla_{\circ}\varepsilon + \nabla_{\circ}m_0\cdot(I-\varepsilon)\right)\cdot\frac{(I-\varepsilon)\cdot m_0}{|F^{-T}\cdot m_0|} + I:\nabla_{\circ}\varepsilon =$$
$$-m_0\cdot\nabla_{\circ}\varepsilon\cdot(I-\varepsilon)\cdot m_0 + \nabla_{\circ}m_0\cdot(I-\varepsilon)\cdot(I-\varepsilon)\cdot m_0 + I:\nabla_{\circ}\varepsilon =$$
$$-m_0\cdot\nabla_{\circ}\varepsilon\cdot I\cdot m_0 + \nabla_{\circ}m_0\cdot(I-2\varepsilon)\cdot m_0 + I:\nabla_{\circ}\varepsilon =$$
$$-\nabla_{\circ}\varepsilon:\left(m_0 \otimes m_0\right) + \nabla_{\circ}m_0\cdot(I-2\varepsilon)\cdot m_0 + I:\nabla_{\circ}\varepsilon = \nabla_{\circ}\varepsilon:\left(I-m_0 \otimes m_0\right) - 2\nabla_{\circ}m_0\cdot\varepsilon\cdot m_0, \quad (129)$$

where we used $\nabla_{\circ}m_0\cdot m_0 = 0$. And for $N$

$$N = \left(\nabla_{\circ}m_0\cdot F^{-1}\cdot F^{-T} + m_0\cdot\nabla_{\circ}F^{-1}\cdot F^{-T} + m_0\cdot F^{-1}\cdot\nabla_{\circ}F^{-T}\right):\frac{I-m_0 \otimes m_0}{|F^{-T}\cdot\nabla_{\circ}\phi|} -$$
$$\frac{\nabla_{\circ}\cdot m_0}{|F^{-T}\cdot\nabla_{\circ}\phi|}\left(m_0\cdot F^{-1}\cdot F^{-T}\cdot m_0\right) -$$
$$\frac{F^{-T}\cdot\nabla_{\circ}\phi}{\left(F^{-T}\cdot\nabla_{\circ}\phi\right)^2}\cdot\left(\nabla_{\circ}F^{-T}\cdot\nabla_{\circ}\phi + F^{-T}\cdot\nabla_{\circ}\nabla_{\circ}\phi\right)\cdot\left(I-m_0 \otimes m_0\right)\cdot F^{-1}\cdot F^{-T}\cdot m_0 \simeq \quad (130)$$

$$\left(\nabla_\circ \boldsymbol{m}_0 \cdot (\boldsymbol{I}-\boldsymbol{\varepsilon})\cdot(\boldsymbol{I}-\boldsymbol{\varepsilon}) - \boldsymbol{m}_0\cdot\nabla_\circ\boldsymbol{\varepsilon}\cdot(\boldsymbol{I}-\boldsymbol{\varepsilon}) - \boldsymbol{m}_0\cdot(\boldsymbol{I}-\boldsymbol{\varepsilon})\cdot\nabla_\circ\boldsymbol{\varepsilon}\right):\frac{\boldsymbol{I}-\boldsymbol{m}_0\otimes\boldsymbol{m}_0}{|\nabla_\circ\phi|} -$$

$$\frac{\nabla_\circ\cdot\boldsymbol{m}_0}{|\nabla_\circ\phi|}\boldsymbol{m}_0\cdot(\boldsymbol{I}-\boldsymbol{\varepsilon})\cdot(\boldsymbol{I}-\boldsymbol{\varepsilon})\cdot\boldsymbol{m}_0 -$$

$$\frac{(\boldsymbol{I}-\boldsymbol{\varepsilon})\cdot\nabla_\circ\phi}{(\nabla_\circ\phi)^2}\cdot\left(-\nabla_\circ\boldsymbol{\varepsilon}\cdot\nabla_\circ\phi + (\boldsymbol{I}-\boldsymbol{\varepsilon})\cdot\nabla_\circ\nabla_\circ\phi\right)\cdot\left(\boldsymbol{I}-\boldsymbol{m}_0\otimes\boldsymbol{m}_0\right)\cdot(\boldsymbol{I}-\boldsymbol{\varepsilon})\cdot(\boldsymbol{I}-\boldsymbol{\varepsilon})\cdot\boldsymbol{m}_0.$$

Since $\left(\boldsymbol{I}-\boldsymbol{m}_0\otimes\boldsymbol{m}_0\right)\cdot(\boldsymbol{I}-\boldsymbol{\varepsilon})\cdot(\boldsymbol{I}-\boldsymbol{\varepsilon})\cdot\boldsymbol{m}_0 \simeq \left(\boldsymbol{I}-\boldsymbol{m}_0\otimes\boldsymbol{m}_0\right)\cdot\boldsymbol{m}_0 = \boldsymbol{0}$, the last term vanishes and

$$\begin{aligned}
N &= \left(\nabla_\circ\boldsymbol{m}_0\cdot(\boldsymbol{I}-2\boldsymbol{\varepsilon}) - 2\boldsymbol{m}_0\cdot\nabla_\circ\boldsymbol{\varepsilon}\right):\frac{\boldsymbol{I}-\boldsymbol{m}_0\otimes\boldsymbol{m}_0}{|\nabla_\circ\phi|} - \\
&\frac{\nabla_\circ\cdot\boldsymbol{m}_0}{|\nabla_\circ\phi|}\boldsymbol{m}_0\cdot(\boldsymbol{I}-2\boldsymbol{\varepsilon})\cdot\boldsymbol{m}_0 = \left(\nabla_\circ\boldsymbol{m}_0 - 2\nabla_\circ\boldsymbol{m}_0\cdot\boldsymbol{\varepsilon} - 2\boldsymbol{m}_0\cdot\nabla_\circ\boldsymbol{\varepsilon}\right):\frac{\boldsymbol{I}-\boldsymbol{m}_0\otimes\boldsymbol{m}_0}{|\nabla_\circ\phi|} - \\
&\frac{\nabla_\circ\cdot\boldsymbol{m}_0}{|\nabla_\circ\phi|}\left(\boldsymbol{m}_0 - 2\boldsymbol{m}_0\cdot\boldsymbol{\varepsilon}\right)\cdot\boldsymbol{m}_0 = -2\nabla_\circ\boldsymbol{m}_0:\left(\boldsymbol{\varepsilon}\cdot\frac{\boldsymbol{I}-\boldsymbol{m}_0\otimes\boldsymbol{m}_0}{|\nabla_\circ\phi|}\right) - \\
&2\boldsymbol{m}_0\cdot\nabla_\circ\boldsymbol{\varepsilon}:\frac{\boldsymbol{I}-\boldsymbol{m}_0\otimes\boldsymbol{m}_0}{|\nabla_\circ\phi|} - 2\boldsymbol{m}_0\cdot\boldsymbol{\varepsilon}\cdot\left(\boldsymbol{m}_0\nabla_\circ^2\phi - \boldsymbol{m}_0\cdot(\nabla_\circ\nabla_\circ\phi)^T\right)/(\nabla_\circ\phi)^2,
\end{aligned} \tag{131}$$

where we used $\nabla_\circ\boldsymbol{m}_0:\left(\boldsymbol{I}-\boldsymbol{m}_0\otimes\boldsymbol{m}_0\right) = \nabla_\circ\cdot\boldsymbol{m}_0 - \boldsymbol{m}_0\cdot\nabla_\circ\boldsymbol{m}_0\cdot\boldsymbol{m}_0 = \nabla_\circ\cdot\boldsymbol{m}_0$ and $\boldsymbol{m}_0\cdot\boldsymbol{m}_0\nabla_\circ\cdot\boldsymbol{m}_0 = \nabla_\circ\cdot\boldsymbol{m}_0$. Eqs (128)-(131) are the same as Eq. (125)-(127) from a different derivation. We did not introduce the rotation tensor into the derivations here because it vanishes, similar to Eq. (124).